\documentclass[sn-nature]{sn-jnl}
 
\usepackage{graphicx}%
\usepackage{multirow}%
\usepackage{amsmath,amssymb,amsfonts}%
\usepackage{amsthm}%
\usepackage{mathrsfs}%
\usepackage[title]{appendix}%
\usepackage{xcolor}%
\usepackage{textcomp}%
\usepackage{manyfoot}%
\usepackage{booktabs}%
\usepackage{algorithm}%
\usepackage{algorithmicx}%
\usepackage{algpseudocode}%
\usepackage{listings}%
\usepackage{subcaption}
\usepackage{chngcntr}  

\newcounter{suppfigure}

\newenvironment{suppfigure}[1][]{%
    \refstepcounter{suppfigure}
    \renewcommand{\thefigure}{S\arabic{suppfigure}}
    \begin{figure}[#1]
}{%
    \end{figure}
}
\newcommand{\suppfigcaption}[1]{%
    \caption{#1}
}

\unnumbered

\begin{document}

\title[]{Scalable and accurate simulation of electrolyte solutions with quantum chemical accuracy.}

\author[1]{\fnm{Junji} \sur{Zhang}}
\author[1]{\fnm{Joshua} \sur{Pagotto}}
\author[2]{\fnm{Tim} \sur{Gould}}
\author*[1,2]{\fnm{Timothy T.} \sur{Duignan}}\email{tim@duignan.net}

\affil[1]{\orgdiv{School of Chemical Engineering}, \orgname{The University of Queensland}, \orgaddress{ \city{Brisbane}, \state{QLD}  \postcode{4072}, \country{Australia}}}

\affil*[2]{\orgdiv{Queensland Micro- and Nanotechnology Centre}, \orgname{Griffith University}, \orgaddress{\city{Nathan}, \state{QLD} \postcode{4111}, \country{Australia}}}

\abstract{
Electrolyte solutions play critical role in a vast range of important applications, yet an accurate and scalable method of predicting their properties without fitting to experiment has remained out of reach, despite over a century of effort. Here, we combine state-of-the-art density functional theory and equivariant neural network potentials to demonstrate this capability, reproducing key structural, thermodynamic, and kinetic properties. We show that neural network potentials (NNPs) can be recursively trained on a subset of their own output to enable coarse-grained/continuum-solvent molecular simulations that can access much longer timescales than possible with all atom simulations. We observe the surprising formation of Li cation dimers along with identical anion-anion pairing of chloride and bromide anions. Finally, we reproduce simulate the crystal phase and infinite dilution pairing free energies despite being trained only on moderate concentration solutions. This approach should be scaled to build a greatly expanded database of electrolyte solution properties than currently exists.}
\maketitle
\maketitle
\section{Introduction}
Molecular-scale processes, occurring at the level of thousands of atoms, are at the heart of chemistry and biology. These processes obey the laws of quantum mechanics and dictate the behavior of a vast range of crucial systems, yet they remain elusive, hidden from direct observation. We rely on indirect experiments and models to piece together their mysteries, but this limits our mastery over vital biological, chemical, and material systems. Unlocking these secrets could revolutionise our understanding and control of the world at its most fundamental level.

Accurate and efficient first principles molecular dynamics simulations (FPMD) of these processes would be a transformatively useful tool for achieving this goal. This would enable the direct observation of key processes; the calculation of important properties such as free energies; and the generation of abundant training data for machine learning models. However, achieving this goal requires the solution to two inter-related problems: how to run calculations for sufficiently long times whilst employing sufficiently accurate models for the quantum mechanical interactions.

A solution to the first problem is now
possible thanks to neural network potentials (NNPs).\cite{deringerOriginsStructuralElectronic2021,Galib2021,kapilFirstprinciplesPhaseDiagram2022,boreRealisticPhaseDiagram2023} This approach trains a neural network to reproduce quantum chemistry calculations, usually generated with density functional theory (DFT).\cite{Mater2019,Noe2020,Behler2021,White2021,Kocer2022,Wen2022,kolluruOpenChallengesDeveloping2022,tokitaHowTrainNeural2023}
NNPs are several orders of magnitude faster than direct calculations and are rapidly improving thanks to advances in the treatment of electrostatics.\cite{yaoTensorMol0ModelChemistry2018,Yue2021,Behler2021,Gao2021a,jacobsonTransferableNeuralNetwork2022} and the use of equivariance,\cite{Satorras2021,Batzner2021,Batatia2022,Tholke2022,Liao2022} which means that when the input to the NNP undergoes a symmetry transformation (like rotation or translation), the network’s output transforms in a predictable and consistent way under the same transformation.

Furthermore, coarse grained NNPs can also be trained and used to run coarse-grained molecular dynamics simulations that ignore irrelevant parts of a systems, enabling additional orders of magnitude acceleration. 
~\cite{husicCoarseGrainingMolecular2020,looseCoarseGrainingEquivariantNeural2023a,majewskiMachineLearningCoarseGrained2023,wilsonAnisotropicMolecularCoarsegraining2023,costeDevelopingImplicitSolvation2023a} 
The simplest example of a coarse grained NNP is a continuum solvent NNP. This model takes as input the positions of all of the solutes in the system and outputs the average forces on them, where the averaging is over solvent configurations at equilibrium. The training data can therefore be obtained from equilibrium all atom MD simulation where the data on the solvent molecules is simply discarded. If the NNP correctly reproduces these average forces then it must be correctly modeling the free energy of the solutes as this corresponds to the `potential of mean force'.  

So far these coarse grained NNPs have only been trained on classical all atom MD simulations, this is because DFT is too expensive to obtain the large datasets required. However, it should be possible to train them on all atom NNP-MD trained on DFT data. 

Electrolyte solutions are a natural initial system to apply these tools to because determining the properties of electrolytes from first principles is a foundational problem of physical chemistry.
Additionally, these solutions play a key role in a vast range of important processes and systems.
Electrolytes are ubiquitous throughout chemical engineering and their equilibrium constants, solubilities, and reaction rates etc are critically important.
The standard way to model these solutions today is with the century old Debye-H\"{u}ckel theory augmented with empirically fitted parameters.  
Bypassing this reliance on empirical data~\cite{VaqueAura2021} is of enormous practical value as obtaining high quality experimental data can be prohibitively challenging for many systems. 


However, studying electrolyte solutions exposes the second problem -- how to achieve sufficient accuracy. Initial work on applying NNPs to this problem has shown significant promise.%
~\cite{Hellstrom2017,Smith2017,yaoTensorMol0ModelChemistry2018,Onat2018,Galib2021,Zhang2021g,Shi2021a,Zhang2022,Gao2021a,zhangMolecularDynamicsSimulations2022,zhangPredictiveDesignElectrolyte2022,Malosso2022,Shi2022a,jacobsonTransferableNeuralNetwork2022,dajnowiczHighDimensionalNeuralNetwork2022,Wang2022,zhangWhyDissolvingSalt2023,avulaUnderstandingAnomalousDiffusion2023a,jacobsonLeveragingMultitaskLearning2023,guoAL4GAPActiveLearning2023a,oneillPairNotPair2024,gongBAMBOOPredictiveTransferable2024}  However, classical forcefields and low level DFT, such as generalised gradient approximation, are not sufficiently accurate%
~\cite{Riera2017,Paesani2019,Wagle2020,Duignan2020}
to predict the complex dynamic interactions that govern electrolyte solution behaviour. Hence, quantitative prediction of important electrolyte solution properties such as activity coefficients, that depend sensitively on ion pairing has yet to be demonstrated. It has been particularly difficult to simulate ion pairing at low concentrations with NNP-MD because these pairs are much rarer than ion-solvent or solvent-solvent interactions and their energies are sensitive to the accuracy of the underlying quantum model, which means it is challenging to generate enough data with a high enough level of theory to train an accurate NNP. 

\begin{figure}
\centering
\includegraphics[width=.9\textwidth]{{{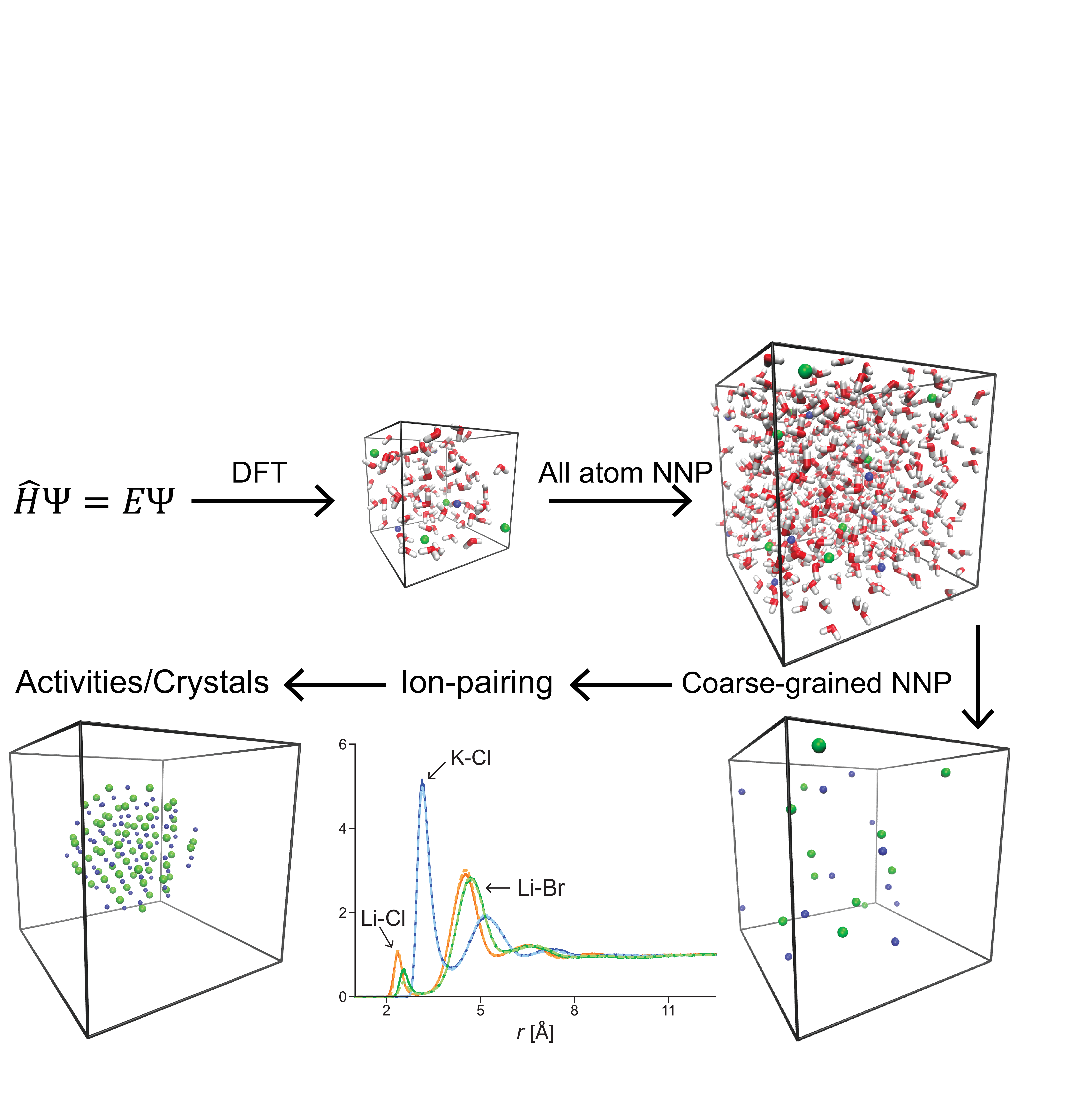}}}
\caption{\textbf{Workflow:}  Small, short first principles molecular dynamics (FPMD) simulations are run with CP2K. Energies/forces computed with DC-r$^2$SCAN are used to train a neural network potential (NNP), which enables much faster simulations with a larger cell, while maintaining high accuracy. Forces and coordinates of the ions alone are output from the NNP-MD and used to train a coarse-grained/continuum-solvent NNP, which enables even faster coarse grained MD simulations to be performed and the simulation of crystals, despite being trained only on the solution-phase.}\label{fig:workflow}
\end{figure}

Density-corrected DFT\cite{simImprovingResultsImproving2022} (DC-DFT) offers a solution to the accuracy problem, as it has been shown to reduce errors, such as delocalization and self-interaction, which cause standard density functional approximations (DFAs) to inaccurately describe ions. DC-r$^2$SCAN, which uses a Hartree-Fock electron density as input into the renormalized strongly constrained and appropriately normalised (SCAN) DFA, has demonstrated significant promise for aqueous ionic systems where it has been carefully validated in comparison to higher levels of quantum chemical theory;%
~\cite{furness2020,Dasgupta2021,palosConsistentDensityFunctional2023a,simImprovingResultsImproving2022,songExtendingDensityFunctional2023,belleflammeRadicalsAqueousSolution2023}
although favorable cancellation of errors may still be playing a role.~\cite{kaplanHowDoesHFDFT2024a}
However, this accuracy comes at the expense of unfavourable computational cost, which hampers modelling of large systems or long time scales.

Here, we demonstrate that DC-DFT can be used to train all atom NNPs to run accurate all-atom molecular dynamics of three electrolyte solutions, with explicit long-range electrostatics described by a continuum-solvent model.
These all atom NNP-MD simulations require minimal training data (hundreds of frames) at a single concentration, allowing us to fit the NNP to accurate (but slow) DC-r$^2$SCAN data, but can be used to simulate at both higher and lower concentration.
We observe the formation of previously unknown Li cation dimers and almost identical anion pairing of chloride and bromide anions. 
We also demonstrate excellent agreement with experimental structural, kinetic, and thermodynamic properties. 

We then demonstrate that the all atom NNP-MD simulations can be used to train coarse-grained/continuum-solvent NNPs where only the ions are included. These coarse grained NNPs can be used to run MD simulations that are dramatically faster,
yet remarkably, they demonstrate complex, out-of-distribution behaviour such as crystal dissolution, and accurate infinite dilution pairing free energies; despite being trained only on moderate concentration solutions.
This work thereby demonstrates the ability of NNPs to predict properties not directly seen in their training data, by exploiting fundamental physics like equivariance together with diverse and accurate quantum mechanical data from DC-r$^2$SCAN.
This workflow is outlined in Figure~\ref{fig:workflow} and fully described in the Computational details section. 


\section{Results and Discussion}

\begin{figure*}
    \begin{subfigure}[]{0.41\textwidth}
        \centering
        \includegraphics[width=\textwidth]{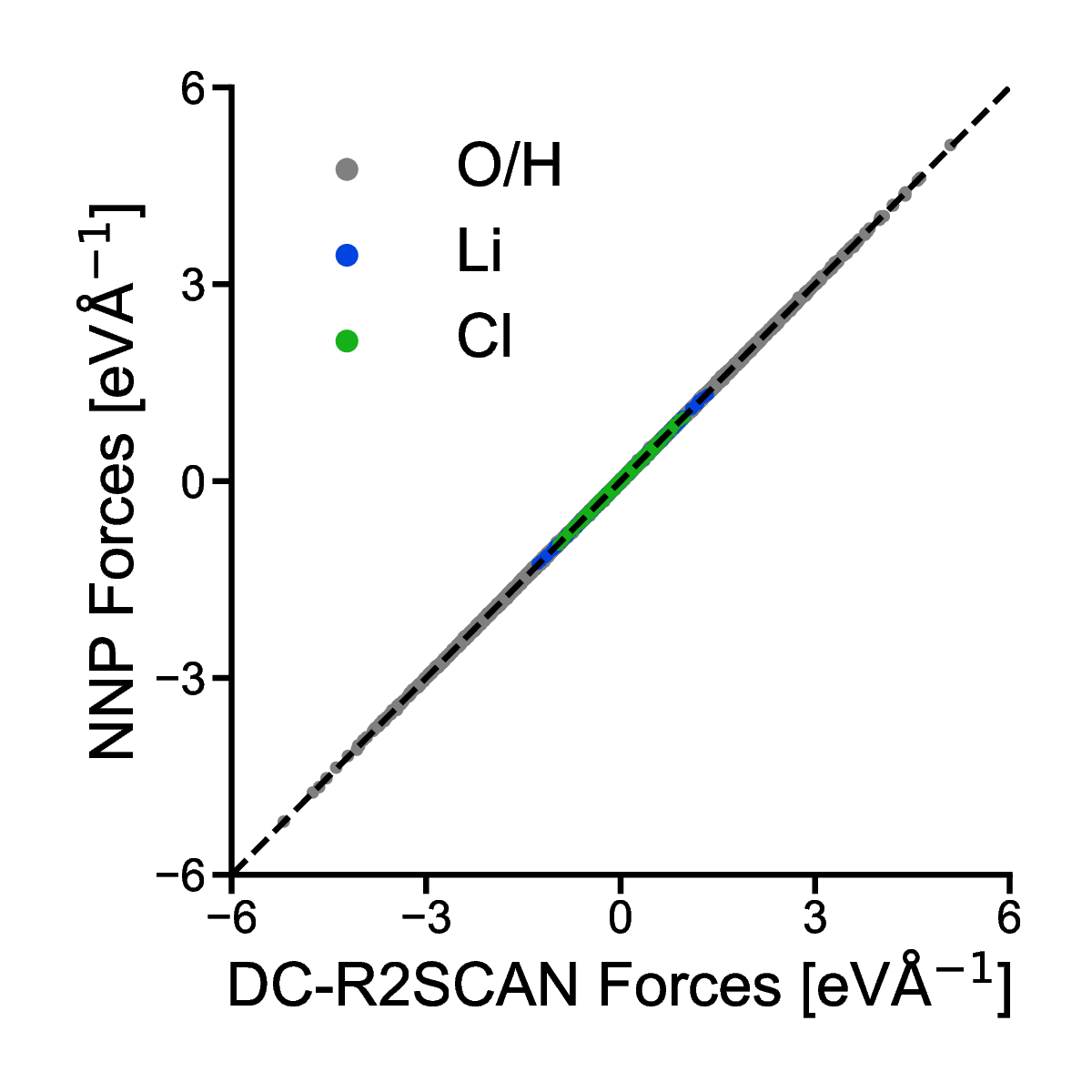}
        \caption{Force comparison}\label{fig:Forcecorre}
    \end{subfigure}
    \hspace{0.07\textwidth}
    \begin{subfigure}[]{0.49\textwidth}
        \includegraphics[width=\textwidth]{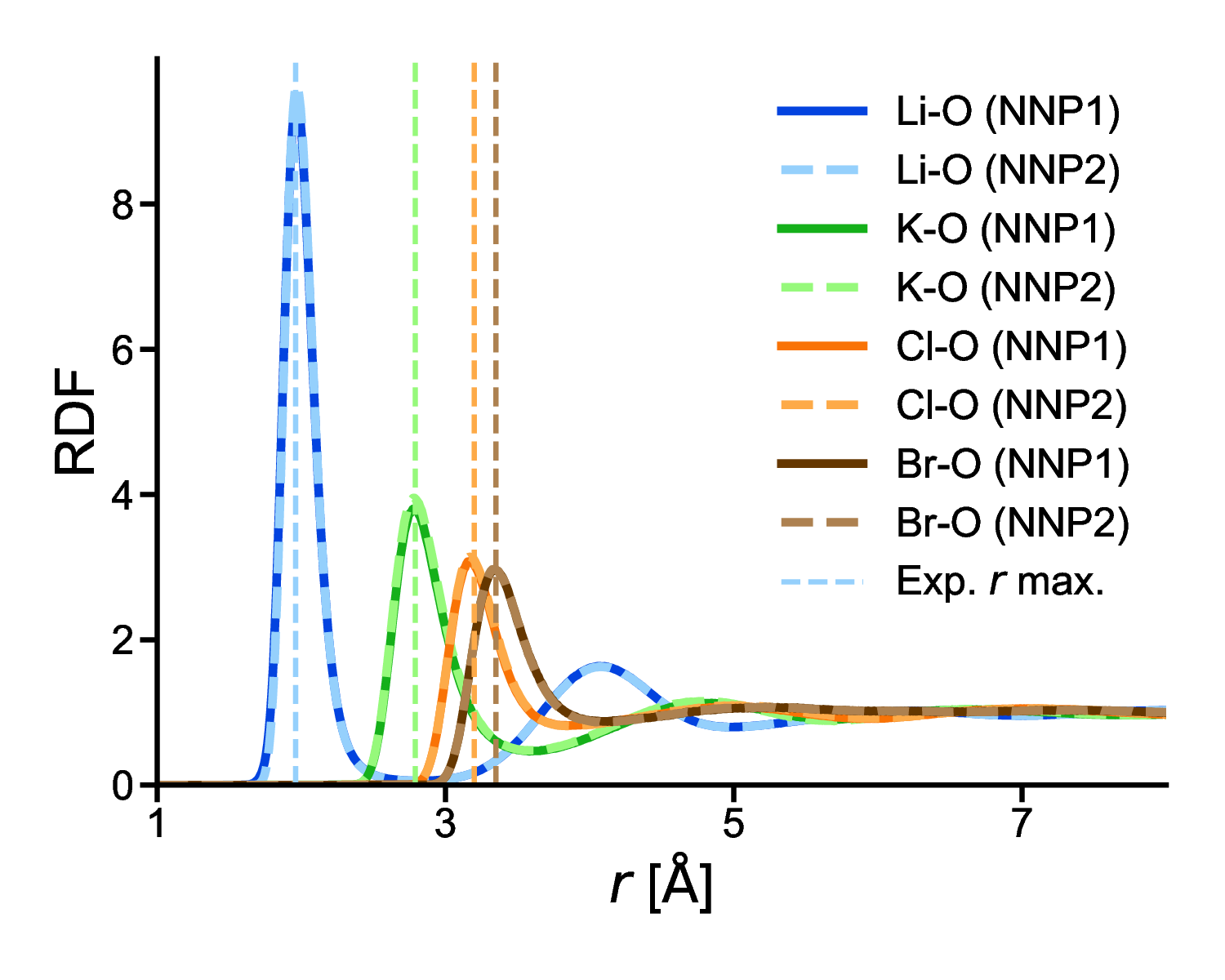}
        \caption{Ion-oxygen RDFs}\label{fig:RDFIonOwatcomp}
    \end{subfigure}
    \\
    \begin{subfigure}[]{0.49\textwidth}
        \includegraphics[width=\textwidth]{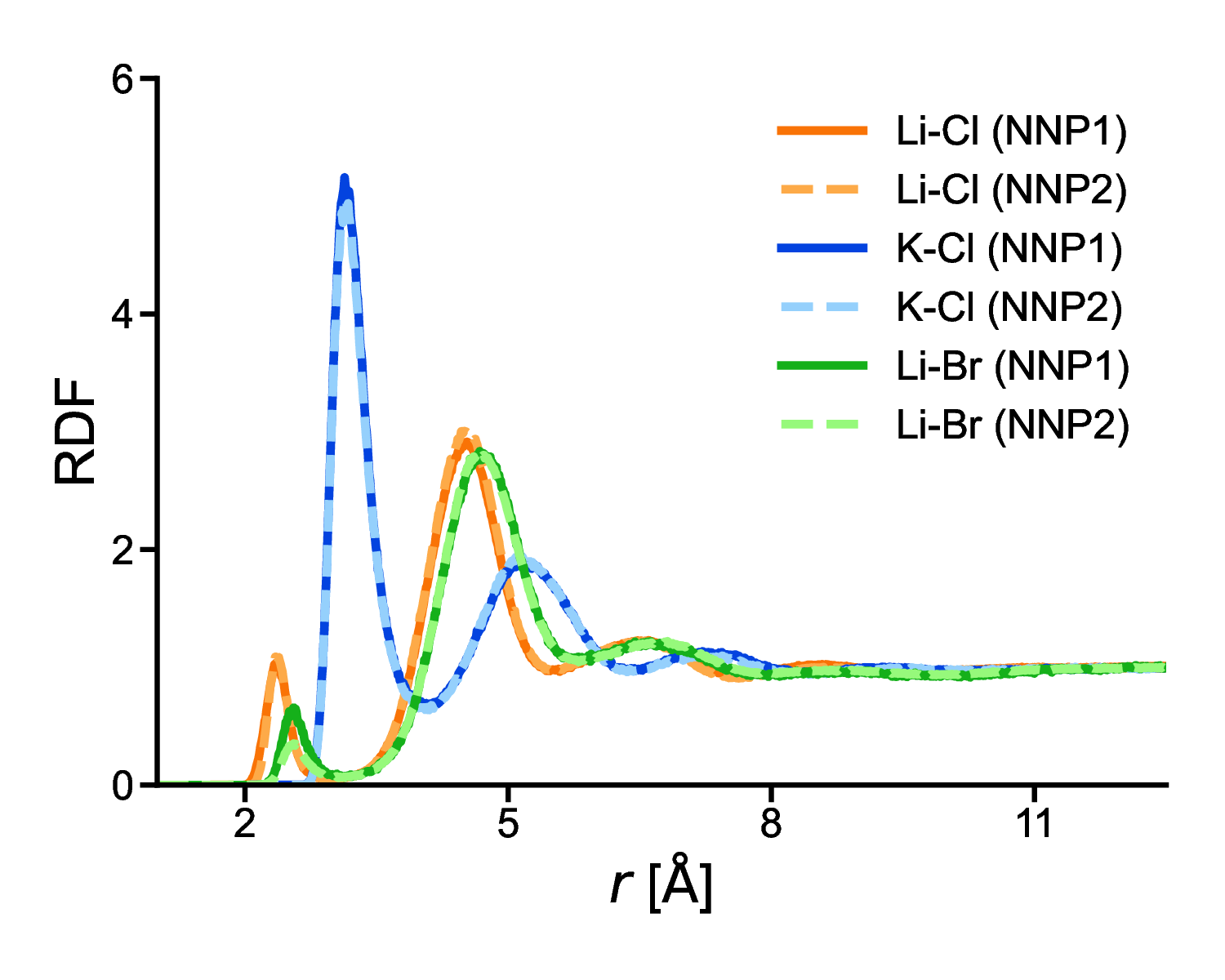}
        \caption{Cation-anion RDFs at 2.5 M}\label{fig:RDFCatAn2p5M}
    \end{subfigure}
    \begin{subfigure}[]{0.49\textwidth}
        \includegraphics[width=\textwidth]{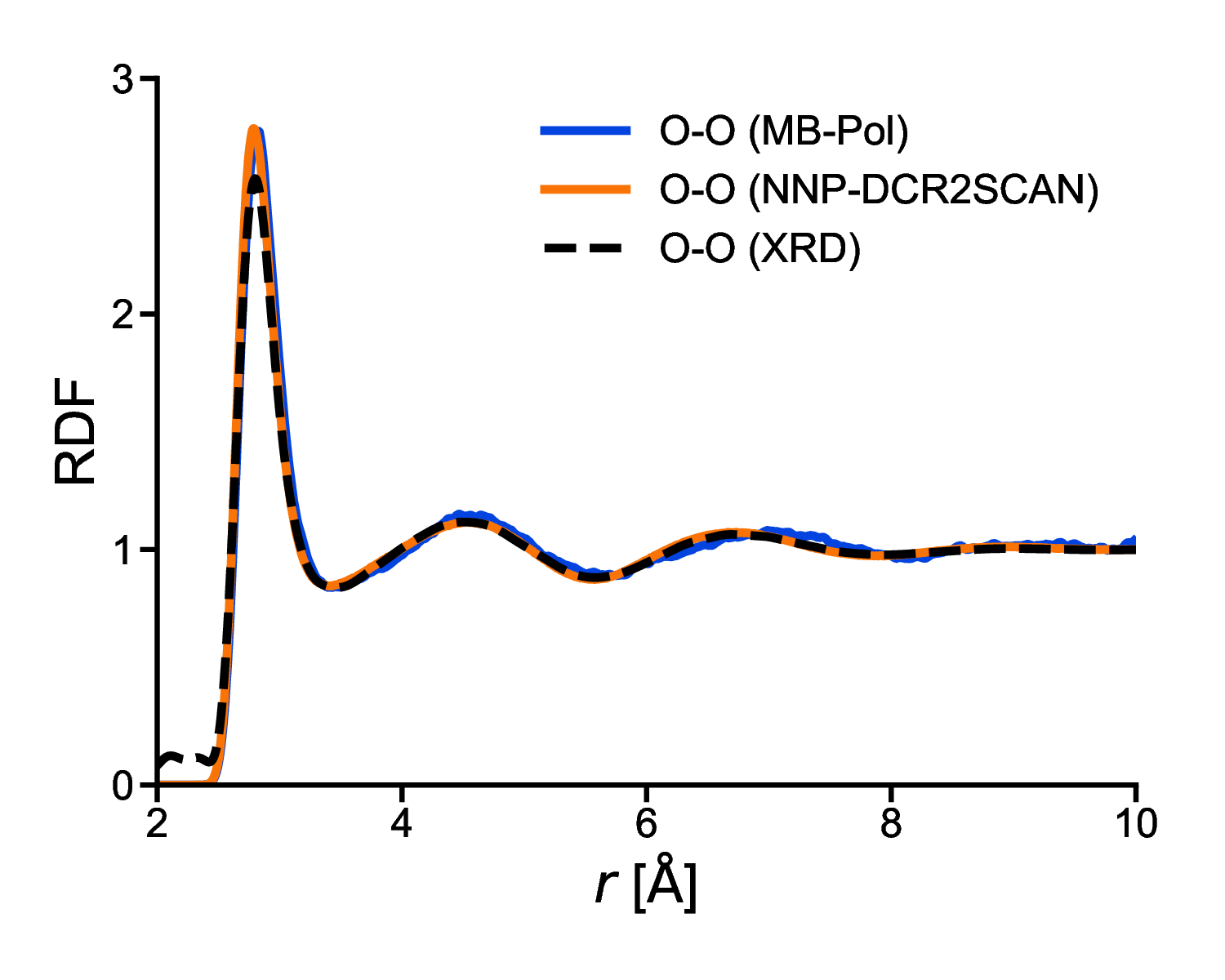}
        \caption{Oxygen-oxygen RDF for pure water.}\label{fig:RDFOO}
    \end{subfigure}
    \suppfigcaption{(a) DFT forces are compared with the predictions of the NNP for LiCl. (b) Comparison of the ion-oxygen RDFs predictions of two NNPs trained with different seeds. Good agreement with neutron and X-ray diffraction (XRD) measurements of ion oxygen RDF is also observed. (c) Comparison of the cation-anion RDFs predictions of two NNPs trained with different initial seeds. (d) Comparison of the predicted O-O RDF from a pure water all atom NNP-MD simulation with experimental XRD measurements\cite{Skinner2013} and the MB-pol water model.\cite{Medders2014}}
    \label{fig:RDFs}
\end{figure*}
The NNP (built with NequIP\cite{Batzner2021}) exploits fundamental symmetries to build highly accurate potentials from strikingly small MD simulation data sets of  500-700 frames made up of 88 molecules each at only a single concentration (2.5 M). 
Its data efficiency enables the model to be trained on highly accurate yet costly DC-r$^2$SCAN DFT calculations carried out with CP2K.~\cite{VandeVondele2005,Kuhne2020,belleflammeRadicalsAqueousSolution2023}
The resulting NNPs provide highly accurate and reproducible prediction of forces, ion-solvent, ion-ion and pure water structure as shown in Figure~\ref{fig:RDFs}.

Two NNPs were generated for each electrolyte using different random seeds, in order to validate them. For LiCl, 200 frames were extracted from the all atom NNP-MD simulation, and forces were recomputed with DFT for validation as shown in Figure~\ref{fig:Forcecorre} giving very accurate estimates the forces during NNP-MD with an RMSE below 10 meV\AA$^{-1}$.

Parallel all atom NNP-MD simulations were run for over a nanosecond each with both NNP1 and NNP2 on a system six times larger than the original, containing 512 water molecules and 48 ions. Accessing this time and spatial scale is entirely infeasible with direct FPMD, yet it is critical for studying ion-ion interactions, which display long-range structure larger than the cell size used for the FPMD. The all atom NNP-MD generates approximately 200 ps per day for the larger system on a single V100 GPU. In contrast the coarse grained simulations can produce 300 ps on a single Intel Xeon Scalable ‘Cascade Lake’ processors. Additionally the coarse graining accelerates the dynamics providing an additional speedup. 

Figure~\ref{fig:RDFIonOwatcomp} and Figure~\ref{fig:RDFCatAn2p5M} compares the ion-solvent and ion-ion RDFs computed with the two separate NNPs, showing excellent agreement and demonstrating convergence and the reproducibility of the method. The ion-oxygen peak positions of 1.99, 2.78, 3.17, and 3.34 \AA\ are in good agreement with neutron/X-ray diffraction measurements of 1.96, 2.79, 3.2, and 3.35 \AA\ for lithium, potassium, chloride, and bromide, respectively.\cite{Marcus2009,Ohtaki1993} The ion-oxygen peaks are also in good agreement with direct first principles simulation.\cite{duignanQuantifyingHydrationStructure2020} We use neutron diffraction measurements only for lithium due to the distorting effect of polarization on XRD measurements of its structure.\cite{Ohtaki1993} The ion-ion RDFs are also reproducible. Figure~\ref{fig:RDFCatAn2p5M} demonstrates the strong specificity of cation-anion pairing.  Experimental data on the ion-ion RDFs is extremely limited due to the challenges associated with extracting this signal from the background solvent-solvent and ion-solvent structure. 

For systems of this size nonphysical artifacts can be observed every few nanoseconds, i.e., the formation of a very close-contact ion-ion pair. This indicates that we are operating at the  minimum limit of training data. These errors could be corrected with various strategies such as additional resampling or active learning for generating new training data or with enhanced sampling or higher temperature sampling. These artifacts do not substantially alter the results of this study.  

Water-water interactions help to determine the ion-ion interactions; it is therefore important to test the NNPs' description of the pure water interactions. To do so we, can run a pure water simulation with the all atom NNP trained on 2.5 M. Remarkably, Figure~\ref{fig:RDFOO} shows essentially perfect agreement with experimental XRD measurements. “The small deviation from experiment  in the first peak height can be attributed to the neglect of nuclear quantum effects (NQEs ), which modify the nuclear dynamics but not the electronic potential energy surface.  We can verify this by comparing it to MB-Pol,\cite{Medders2014} a water model which does reproduce the correct peak height when NQEs are included, but which gives very similar results to our model when they are neglected, suggesting that including NQEs in our simulation should also reduce the peak height. Excellent agreement between MB-pol and our model for the OH and HH RDFs is also shown in Figure~\ref{fig:OHHHRDFs}. The fact that, without any data on pure water, we arrive at a model that gives such accurate structural predictions is particularly promising and demonstrates the generalisability of this approach.

\begin{figure*}
\begin{subfigure}[]{0.95\textwidth}
\includegraphics[width=1\textwidth]{{{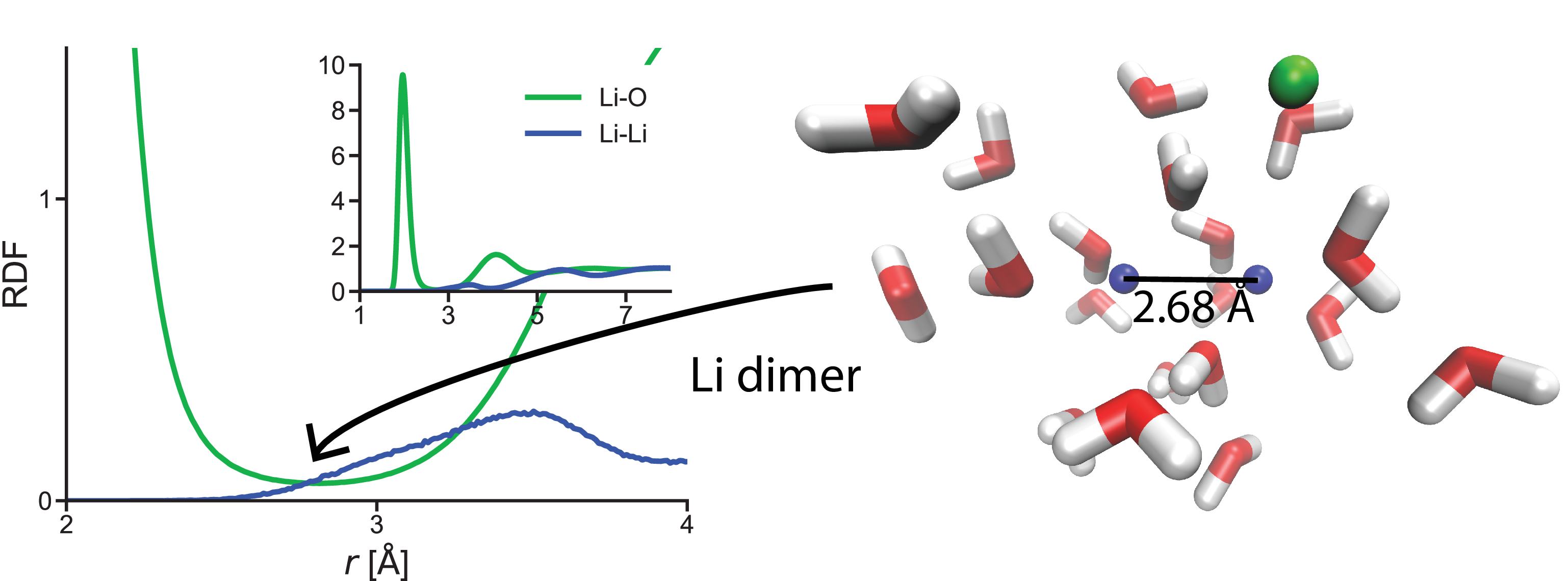}}}
\caption{Lithium-oxygen and lithium-lithium RDFs showing the lithium dimer with an MD snapshot. The zoomed out full RDFs are shown in the inset. 
}\label{fig:RDFLiLiLiO}
\end{subfigure}
\begin{subfigure}[]{0.49\textwidth}
      	 \includegraphics[width=\textwidth]{{{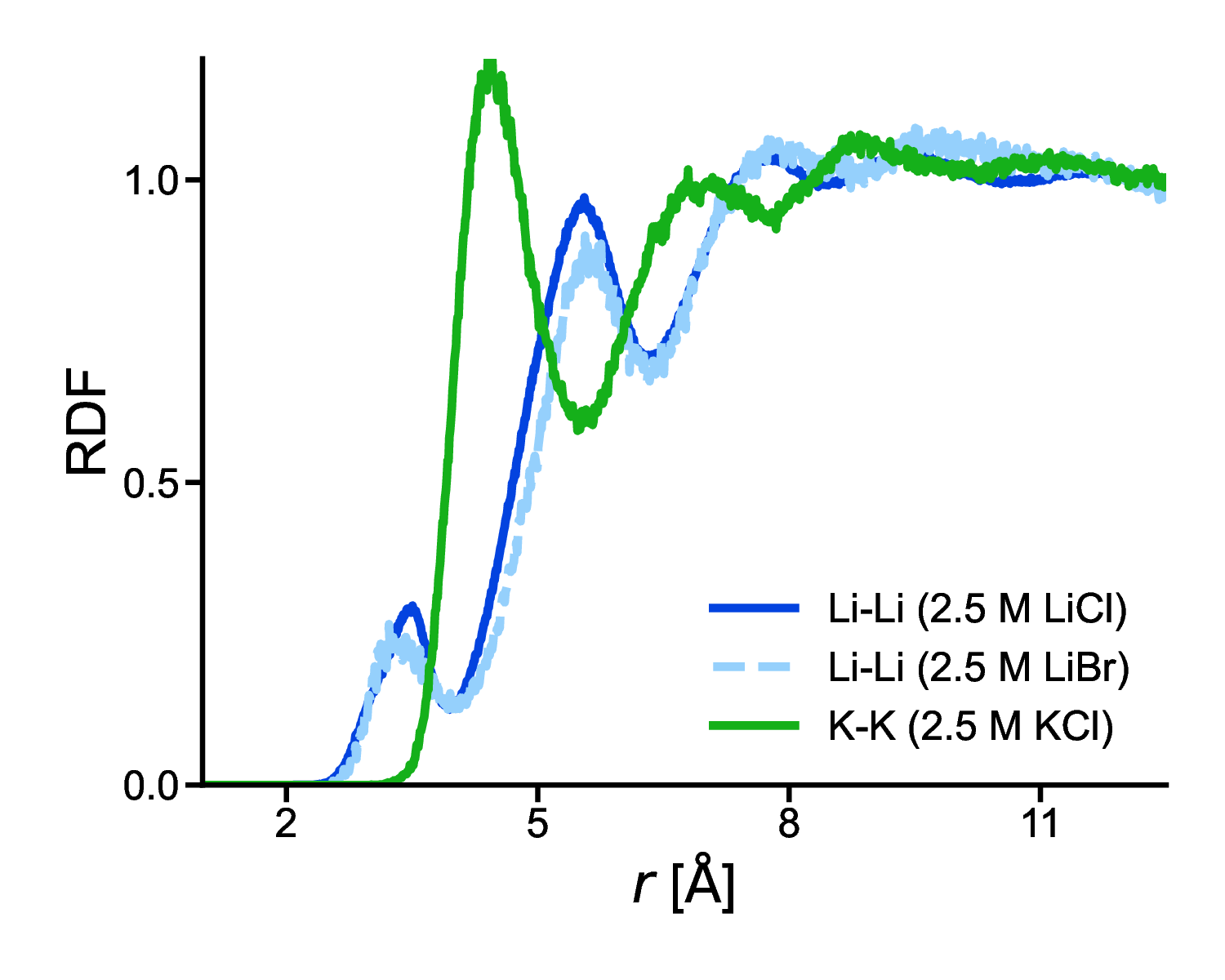}}}
        \caption[]{Cation-cation RDFs at 2.5 M }
        \label{fig:RDFLiLiLiClvsKKClvsLiLiBr}
\end{subfigure}
\begin{subfigure}[]{0.49\textwidth}
      	 \includegraphics[width=\textwidth]{{{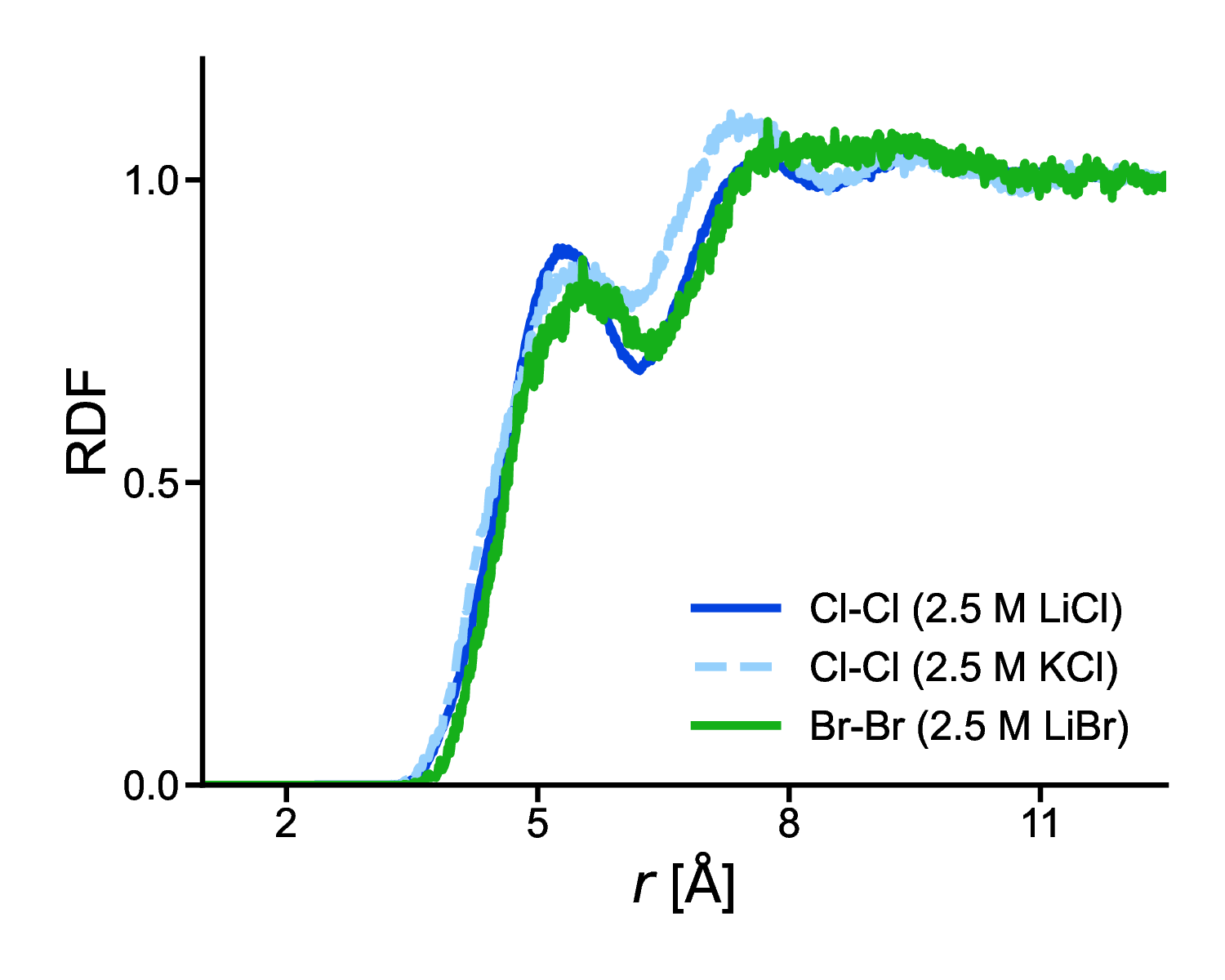}}}
        \caption[]{Anion-Anion RDFs at 2.5 M }
        \label{fig:RDFClClLiClvsKClvsBrBr}
\end{subfigure}
\caption{Cation-cation and anion-anions RDFs revealing the formation of lithium cation dimers and the almost identical pairing of bromide and chloride anions.}
\end{figure*}

\subsection{Lithium dimer formation}
These all atom NNP-MD simulations immediately enable surprising new observations. Firstly, we observed the formation of lithium cation dimers, as shown in Figure~\ref{fig:RDFLiLiLiO}, where one lithium penetrates into the first solvation layer around another lithium ion. The first solvation layer is defined by the first minimum in the Li-O radial distribution function (RDF), which is 2.8 \AA. The smallest separation of the lithium ions observed in the all atom NNP-MD simulations (2.68 \AA) is very similar to the separation of neutral covalent dilithium (2.67 \AA) and is much smaller than the Li-Li distance in LiCl crystal (3.62 \AA). The formation of this species is particularly surprising, given that lithium is an archetypal strongly hydrated ion. Physically, this was believed to correspond to the formation of a tightly bonded first solvation layer of water molecules, which was thought to be impenetrable to other ions.\cite{Collins1997}

We can compute the coordination number of Li-Li coordination nubmer by integrating the RDF up to 4 \AA\, which is distance to the first minima of the Li-Li RDF. This gives a value of 0.05 at 2.4 M, indicating that there is a 5 \% chance of finding a lithium ion dimer at 2.4 M. 

This counter intuitive finding may have important implications for  many biological and chemical systems where lithium plays a critical role.\cite{chenAtomicInsightsFundamental2020} The transient nature of this pair means that this species could not feasibly be identified with direct FPMD simulation, whereas CMD simulations show no indication of it.\cite{Kalcher2009} This is likely attributable to the neglect of charge transfer and polarization effects, which significantly mitigate the electrostatic repulsion, by transferring charge to the surrounding water molecules. Substantial charge transfer is a common observation of ions in water.\cite{Sellner2013} We have confirmed this effect is not an artifact of the all atom NNP-MD simulations by comparing the with DFT calculations on the lithium dimer structure (Figure~\ref{fig:ForcecorrelLidimer}). Figure~\ref{fig:RDFLiLiLiClvsKKClvsLiLiBr} also demonstrates that Li pairing occurs similarly in LiBr electrolyte, indicating the anion is unlikely to play a critical role.

\subsection{Water structure dominated anion pairing}
A second surprising new finding is that the pairing of bromide anions in solution appears to be almost identical to that of chloride anions, as shown in Figure~\ref{fig:RDFClClLiClvsKClvsBrBr}. This indicates that the pairing of anions in water is strongly determined by the surrounding water structure rather than by the inherent size of the anions. This is likely related to the water-bridging structure formed by a water molecule that hydrogen bonds with both anions simultaneously. This behavior is also totally different from the cation-cation pairing, where no bridging occurs and a significant difference between cations is observed, as shown in Figure~\ref{fig:RDFLiLiLiClvsKKClvsLiLiBr}. 

\subsection{Ion pairing: prediction and validation}
The reliability of the ion-ion RDFs can be validated by computing thermodynamic properties. Specifically, the activity coefficients as a function of concentration have been shown to be highly sensitive to the strength of ion-ion interactions. The accurate prediction of these quantities at very low concentration provided the original validation of  Debye-Hückel theory.

\begin{figure} [!htbp]
  \begin{subfigure}[]{0.48\textwidth}
      	 \includegraphics[width=\textwidth]{{{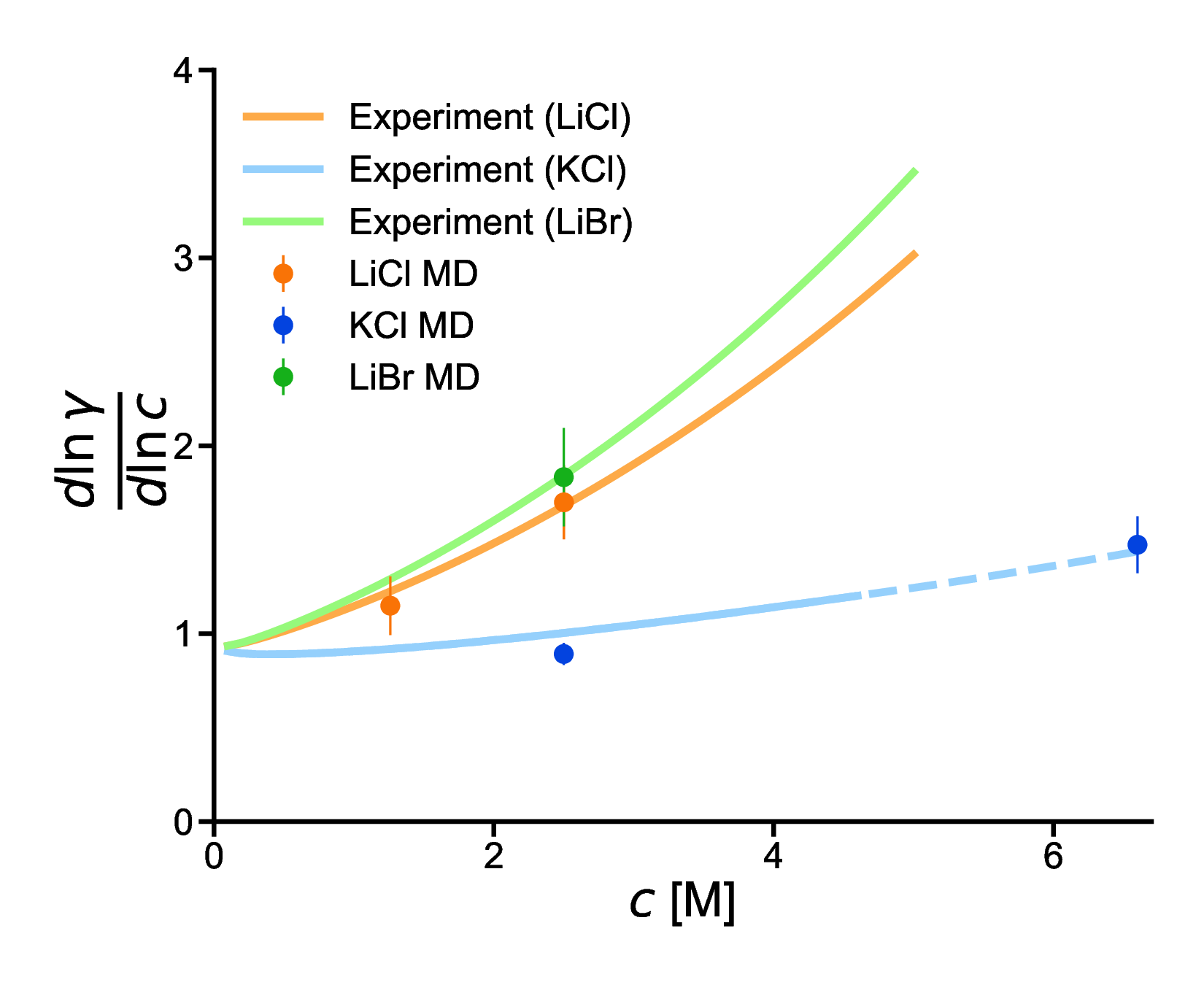}}}
        \caption[]{Activity coefficient derivatives.}
        \label{fig:KBactcomp}
\end{subfigure}
\begin{subfigure}[]{0.48\textwidth}
      	 \includegraphics[width=\textwidth]{{{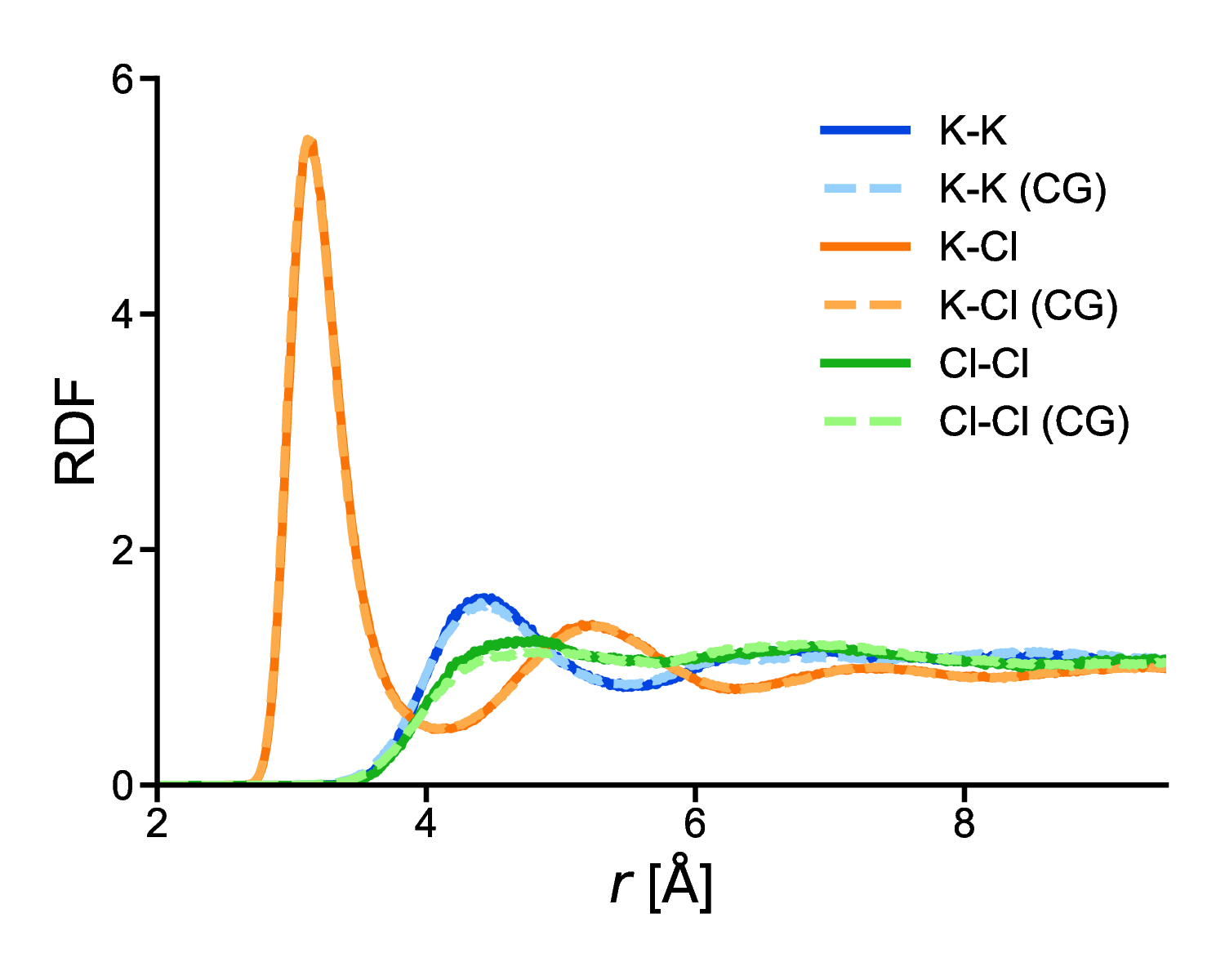}}}
        \caption[]{KCl at 6.6M.}
        \label{fig:RDFKCll36p6M-cg1comp}
\end{subfigure}\\
  \begin{subfigure}[]{0.48\textwidth}
      	 \includegraphics[width=\textwidth]{{{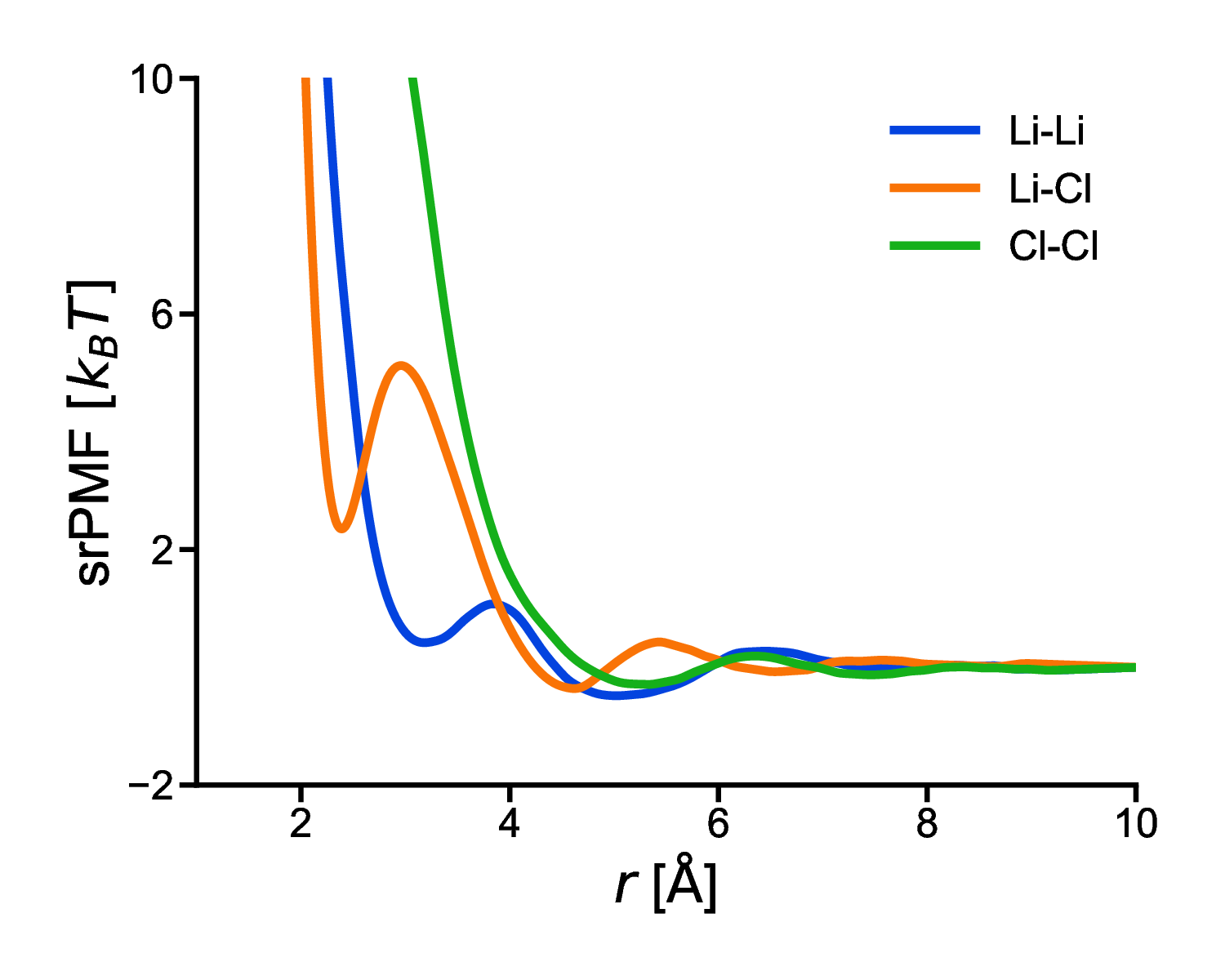}}}
        \caption[]{LiCl PMFs computed from the CG model.}
        \label{fig:LiCl2BSRPMF}
\end{subfigure}
\begin{subfigure}[]{0.48\textwidth}
      	 \includegraphics[width=\textwidth]{{{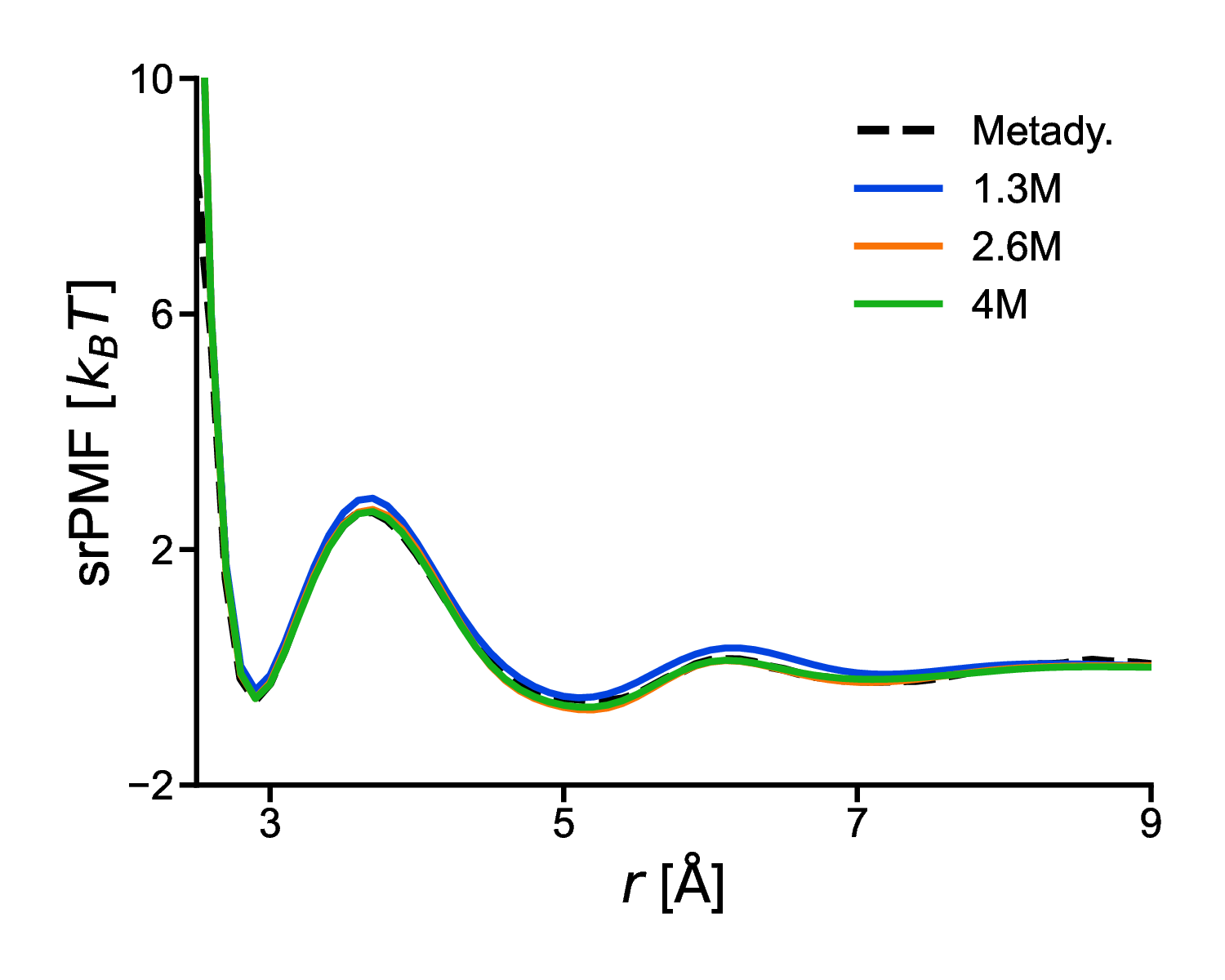}}}
        \caption[]{Classical NaCl short range PMFs.}
        \label{fig:NaNasrPMFmetavsNequIp}
\end{subfigure}
\caption{(a) Comparison of experimental activity coefficient derivatives with predictions using Kirkwood-Buff theory and RDFs from the all atom NNP-MD simulations. (b) Comparison of all-atom RDFs with coarse-grained RDFs for KCl at 6.6 M. (c)  Infinite dilution PMFs for LiCl computed with the coarse-grained model trained at 2.5 M using DFT data. (d) Short ranged infinite dilution potentials of mean force (srPMFs), i.e., free energies, for the NaCl classical force-field computed with the coarse-grained model at various concentrations compared with direct calculation using enhanced sampling. The long range Coulomb component has been removed and thermal units on the y axis are used for (c) and (d). Metadyn. refers to metadynamics, a direct method of computing PMFs.}
\end{figure}

Here, we use Kirkwood-Buff theory to compute a derivative of the log activities, achieving good experimental agreement as shown in Figure~\ref{fig:KBactcomp}. The stronger pairing of KCl compared to LiCl is well reproduced. This is a classic example of a counterintuitive `specific ion effect' induced by the solvent, which is reversed compared with the vacuum pairing strength. In contrast, the stronger pairing of LiCl compared with LiBr does follow the expected behaviour in vacuum, and is also well reproduced. Accurately reproducing these binding strengths is key to understanding many important phenomena such as the effect of ions on protein stability, i.e., the so-called `Hofmeister effect'.\cite{Gregory2022a}


We have also predicted the activity coefficient derivatives at 1.3 M for LiCl and 6.6 M for KCl to demonstrate the generalisability to different concentrations not included in the training data. Activity coefficients describe the non-ideal free energy of ions in solution and are critically important for a wide range of applications. Reproducing these requires correctly reproducing the change in Debye screening length as a function of concentration. (See Figure~\ref{fig:cgcomp}). Debye screening refers to the fact that electrostatic interactions become increasingly short range in electrolytes as the concentration increases due to the cloud of counter charge around an ion. The fact that NNP-MD can reproduce this effect, despite using training data from a single concentration (2.5 M) indicates that the NNP is learning a correct representation of the water-water interactions.  That it is accurate at higher concentrations (6.6 M) is particularly surprising given that higher-order many-body effects are likely to occur here that have not been observed at the lower concentrations.
Note that the KCl experimental values rely on extrapolations of the experimental data, as this is above the experimental solubility point. (Figure~\ref{fig:KBactcomp})  This highlights the capability of this method to obtain experimentally inaccessible data.

This demonstrates the capability to generalise to higher concentration.  This constitutes further evidence of the correct physical interactions are being learnt. 

Diffusivities of Li$^+$ and Cl$^-$ ions, as well as water molecules, are also in reasonable agreement with experiment, as shown in Figure~\ref{fig:Diffs}. This is particularly impressive as kinetic properties, such as diffusivities, depend on accurate barrier heights in the potential energy surface. Structures at the barrier heights  will be underrepresented in the training dataset as they are extracted from equilibrium molecular dynamics. 

\subsection{Coarse-graining: prediction and validation}
The all-atom NNP-MD is much faster than FPMD and capable of simulating experimentally relevant timescales, but the computational cost is still non-trivial and more expensive than most classical molecular dynamics (CMD) approaches. To further lower the computational cost of our method, we build a coarse-grained/continuum-solvent NNP of the electrolyte solutions. Specifically, we integrate out the solvent degrees of freedom, resulting in a continuum or implicit solvent model. To do this, we train an coarge grained NNP to learn the forces on the ions averaged over the equilibrium water configurations. The training data comes all atom MD simulations (either classical or NNP). This corresponds to learning a free energy surface, also referred to as a potential of mean force. Learning free energies is a very similar task to learning the full all-atom potential energy surface and thus also benefits from equivariance, as has recently been demonstrated for pure water.\cite{looseCoarseGrainingEquivariantNeural2023a} As before, we compute the long-range electrostatic interactions separately using Coulomb's law.  The coarse-grained MD can accurately reproduce the RDFs from the all-atom MD, as shown by the close match between dashed and solid lines in Figure~\ref{fig:RDFKCll36p6M-cg1comp}. Good agreement is also observed for LiCl (see Figure~\ref{fig:cgcomp}). 

The coarse-grained NNP-MD is much faster than the all-atom NNP, producing fully converged RDFs within a day on a single CPU, this is due to the faster inference time (300 ps per day on a single CPU vs 200 ps per day on a GPU) and due to the fact that coarse-graining accelerates the dynamics. This hierarchical layering of NNPs where a coarse-grained NNP is trained on data from all-atom NNP-MD is promising as a general approach to the long-standing challenge of connecting scales in molecular simulation given its simplicity. \cite{wangMachineLearningCoarseGrained2019,eMachineLearningassistedMultiscale2023} A comparison of the accuracy of the coarse grained and all atom models is difficult as the ground truth PMFs that the coarse grained model are predicting are unknown. It has much higher loss than the all-atom model, but this is due to the inherent noise from the neglect of solvent degrees of freedom, rather than inherent inaccuracy in the model. 

\subsection{Infinite dilution}
To further validate the reliability of this approach, we also trained a coarse-grained/continuum-solvent NNP on a classical all atom force field for NaCl and demonstrated that RDFs can be reproduced at different concentrations reliably, as shown in Figure~\ref{fig:RDFCMDvsCGMD}. We did not train on first principles data for NaCl, as the standard pseudopotential introduces some noise in the forces on the Na ion. This can be corrected with the  Gaussian and Augmented Plane Wave (GAPW) method,\cite{oneillPairNotPair2024} but this has not been implemented with DC-DFT yet. Additionally, classical force fields for NaCl have already been carefully parameterised to experiment and can do a reasonable job computing properties such as activity coefficients.\cite{Weerasinghe2003}

One key property of solutions that has, to date, been difficult to compute is the infinite dilution potential of mean force (PMF), also called the pair potential.  This corresponds to the free energy change as two solutes move apart in water at infinite dilution, i.e., the infinite dilution pairing free energy. Previously, computation of this quantity required enhanced sampling techniques, such as umbrella sampling with large box sizes. However, our method allows us to trivially extract this quantity from the coarse-grained NNP by simply computing the interaction energy between the ion pairs in isolation. These infinite dilution PMFs are shown in Figure~\ref{fig:LiCl2BSRPMF} for LiCl, with the long-range Coulomb term removed. We validate the reliability of this procedure by showing that the PMF calculated in this way reproduces the ground truth infinite dilution PMF for this classical force field as shown in Figure~\ref{fig:NaNasrPMFmetavsNequIp}. This was computed using metadynamics and thermodynamic integration. See Ref~\citenum{} for details.  This indicates that coarse-grained/continuum-solvent NNPs are a promising tool for the rapid calculation of intrinsic binding free energies of species in solution a problem of ubiquitous importance throughout molecular science.

\subsection{Crystallisation}
\begin{figure*}
  \begin{subfigure}[]{0.99\textwidth}
      	 \includegraphics[width=\textwidth]{{{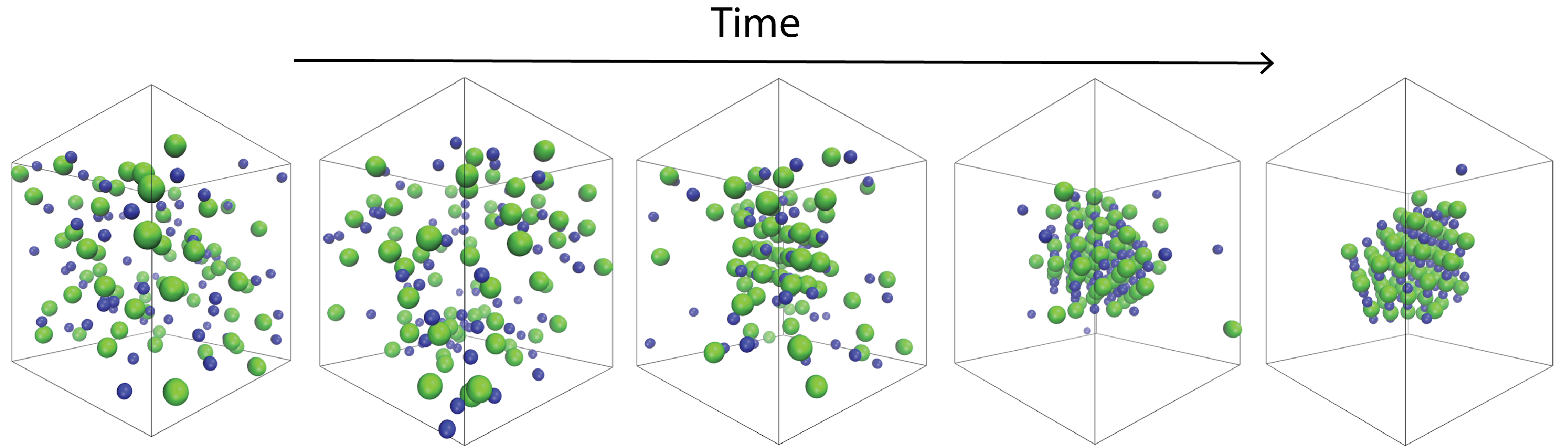}}}
        \caption[]{NaCl nucleation.}
        \label{fig:NaClcryst}
\end{subfigure}\\
\vskip .5cm
\begin{subfigure}[]{0.99\textwidth}
      	 \includegraphics[width=\textwidth]{{{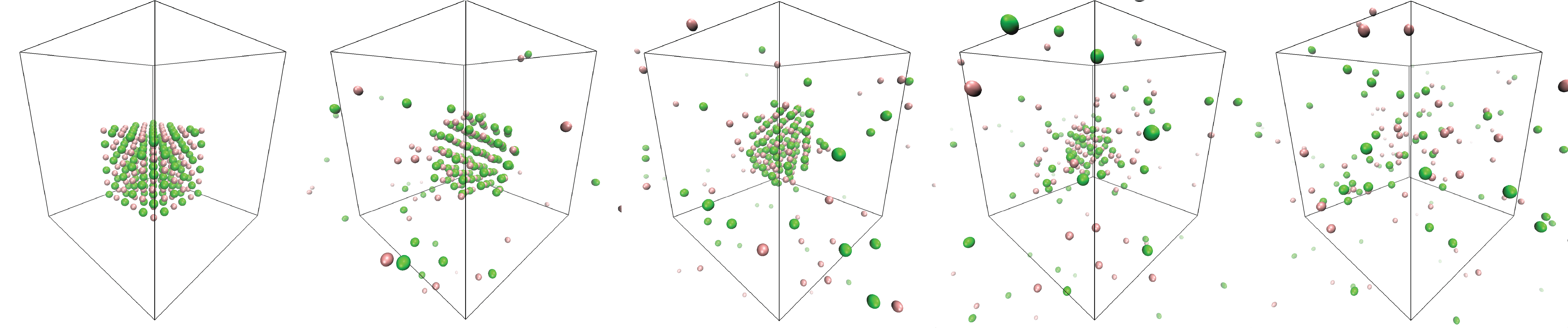}}}
        \caption[]{KCl dissolution.}
        \label{fig:KCldissolution}
\end{subfigure}
\caption{(a) NaCl nucleation and crystal growth observed with a coarse-grained NNP-MD simulation trained on solution-phase classical all atom MD simulation. (b) KCl dissolution with a coarse-grained NNP-MD simulation trained on an all atom NNP-MD simulation, which was trained on DFT data. The snapshots are evenly spaced.}
\end{figure*} 

Most remarkably, while running the coarse-grained NNP trained on the classical NaCl force field, we observed stable crystal nucleation and growth (Fig.~\ref{fig:NaClcryst}), despite the fact that this model was only trained on solution-phase simulations. While we never observed this behaviour in the all-atom MD despite long simulations, this is actually the physically correct behaviour as the solubility predicted by this classical force field is relatively low.\cite{nezbedaRecentProgressMolecular2016} The lattice spacing, i.e., the edge length of the unit cell,  is too large at 6.21 \AA\  compared to the experimental value of 5.63 \AA\, but the correct FCC-structured crystal is formed. This indicates that it should be possible to study complex phenomena, such as crystal nucleation and growth, with a coarse-grained/continuum-solvent NNP. 

We did not observe spontaneous nucleation with the coarse-grained  KCl/LiCl NNP, likely due to their much higher solubility. However,  we did initialise a simulation with a KCl crystal and observed continuous slow dissolution into the liquid starting from the corners while generally maintaining the correct crystalline structure, including the correct lattice parameter of 6.3 \AA. This was followed by a final rapid dissolution step (Fig.~\ref{fig:KCldissolution}). Such behaviour has also recently been observed for NaCl in all-atom NNP-MD simulation.\cite{oneillCrumblingCrystalsDissolution2024} The significant difference, in this case, is that our coarse grained NNP simulations are run on a single CPU within a day with no requirements for crystal-phase training data. 

We also simulated a slab of KCl crystal in periodic boundary conditions with a vacuum region for 1 ns. The vacuum region quickly become saturated with liquid phase KCl ions after which the stable crystal structure of KCl was maintained, indicating that the crystal was not inherently unstable. A movie of the KCl crystal dissolution and NaCl nucleation is available online. 

The use of the output of the all atom neural network potential simulations to train the coarse grain simulations may cause concerns around model collapse, a known issue with training on synthetic data. However, this is not a concern in our case as we are not training on the raw output of a model but instead on molecular simulation data generated with a conservative force field. The coarse graining is therefore a mathematically rigorous process and unlikely to be subject to the same issues. The extent to which this process can be repeated to access even larger scales however, remains an open interesting question. 

\section{Conclusion and future work}
In summary, we have demonstrated the accurate prediction of the structural, kinetic, and thermodynamic properties of aqueous LiCl, KCl, and LiBr electrolyte solutions. The prediction of these properties, particularly activity coefficients, has been a key goal of physical chemistry for over a century. In the process of doing so, we also discovered the formation of lithium dimers and nearly identical anion-anion interactions of chloride and bromide.

In addition, we have demonstrated the ability to recursively train an NNP on its own output to build coarse-grained/continuum-solvent NNPs capable of reproducing all-atom RDFs with further reduced computational demand. These coarse-grained models are capable of reproducing crystal phase behaviour as well as infinite dilution pairing free energies despite being trained only on moderate concentration solution.

Importantly, our approach  combines equivariant NNPs and DC-DFT to achieve this predictive ability with reasonable training data requirements and moderate computational resources.  This means it is feasible to scale this approach to many different electrolyte solutions, enabling the construction of a database of properties of electrolytes across a much wider range of conditions and compositions than currently exists, a task of critical industrial importance. A key focus should be on important electrolytes and properties that are particularly difficult to characterise experimentally, as well as high-temperature and high-pressure conditions, where experimental data is difficult to obtain. 
 


\section{Data availability}
All input scripts and analysis code can be found at: \href{https://github.com/timduignan/Scalable-Electrolyte-Simulation/}{github.com/timduignan/Scalable-Electrolyte-Simulation/}.
Videos of the crystal nucleation and dissolution process can be found at: \href{https://youtube.com/shorts/4ixfnrc-XDg}{youtube.com/shorts/4ixfnrc-XDg} and \href{https://youtube.com/watch?v=eAuS4hDXQBo}{youtube.com/watch?v=eAuS4hDXQBo}.

\section{Computational Details}\label{secA1}
\subsection{FPMD (CP2K)}
We used Born-Oppenheimer first principles molecular dynamics (FPMD) simulations within a constant volume NVT (300~K) ensemble with periodic boundary conditions. The CP2K simulation suite, containing the
\emph{QuickStep} module for the DC-DFT calculations~\cite{VandeVondele2005,Kuhne2020}, was used with a 0.5 fs time step. We used a double $\zeta$  basis set that has been optimized for the condensed phase\cite{VandeVondele2007} in conjunction with GTH pseudopotentials~\cite{Goedecker1996}  optimised for SCAN\cite{Sun2015a,Hutter2021a} and a 1200~Ry cutoff.\cite{Miceli2016,Yao2018} A slightly smaller basis set (DZVP-MOLOPT-SCAN-GTH) was used for LiBr to test the sensitivity of the lithium ion pairing to the basis set size. 
A Nos\'e-Hoover thermostat was attached to every degree of freedom to ensure equilibration.~\cite{Martyna1992} 

An  $\approx$ 10~ps FPMD simulation was run, consisting of 4 cations ions, 4 anions ions, and 80 water molecules for each electrolyte. Cells with fixed dimensions of 13.9$^3$ \AA$^3$, 14.0$^3$ \AA$^3$, and 13.7 $^3$ \AA$^3$ cell were used for LiCl, KCl and LiBr respectively, corresponding to an electrolyte concentration of 2.5 M. The cell size was adjusted to match the experimental density at this concentration. The initial FPMD simulation used the r$^2$SCAN DFA.\cite{furness2020} 

The forces were then computed on samples extracted from FPMD and NNP-MD simulations using the density-corrected r$^2$SCAN level of theory. \cite{simImprovingResultsImproving2022,palosConsistentDensityFunctional2023a} This method has recently been implemented in CP2K.\cite{belleflammeRadicalsAqueousSolution2023} The auxiliary density matrix method (ADMM) was employed to improve the scaling of the four-center two-electron integrals.\cite{guidonAuxiliaryDensityMatrix2010} The Schwarz integral screening threshold was set to 10$^{-5}$ units. A contracted auxiliary basis set (cFIT3) was used to construct the auxiliary density matrix.

\subsection{NNP fitting (NequIP)}
The NequIP package was used for the NNPs.\cite{Batzner2021}
The training dataset was generated as follows: For LiCl, KCl, and LiBr, 458, 492, and 322 frames were evenly sampled directly from an $\approx$ 10 ps FPMD run using r$^2$SCAN. These datasets were used to generate $\approx$  1 nanosecond of all atom NNP-MD data at the FPMD box size. 197, 207, and 201 frames were then evenly sampled from the 1 ns NNP-MD run and force and energies were computed with DC-DFT and added to the training data to improve the diversity for the a second NNP, resulting in a total of 655, 699, and 523 frames for LiCl, KCl and LiBr respectively. 

As our NNP  only has access to local information (short range, $<5$~\AA\cite{Yue2021} )the long-range electrostatic ion-ion interactions were removed from the forces and energies prior to training. These were computed using a dielectrically screened Coulomb interaction.\cite{Pagotto2022} They were then added back in during both all atom and coarse grained NNP-MD simulations. These were calculated with LAMMPS by placing appropriately screened charges on the ions to reproduce dielectric screening and were computed with the particle-particle particle-mesh method.\cite{Hockney1988} Charges of $+1$ for the cations and $-1$ for the anions were used. For KCl and LiBr, a lower dielectric of 50 was used to better describe the high concentration regime to account for the decrease with ion concentration. Whereas for LiCl 78.3 was used, however, we tested LiCl with a dielectric constant of 50 and confirmed that there was no significant difference at 2.5 M. 

The same hyper-parameters were used for the all-atom NNPs. More specifically, 100:1 weighting on forces vs. energies was used in the default loss function.\cite{Batzner2021} We decreased the initial learning rate of 0.01 by a decay factor of 0.5 whenever the validation RMSE in the forces did not see an improvement for five epochs. Training was stopped when the learning rate became smaller than 10$^{-5}$. The model with the lowest validation error was used for simulations. A radial cutoff distance of 5 \AA\ was used. Three layers of interaction blocks were used with the maximum $l$ set to 2, each with 16 features. Only even parity was used. Invariant neurons for the radial network was set to 32. All the other parameters were set to the defaults. An 80:20 training-validation split was used throughout. The validation loss generally appears higher than the training loss perhaps indicating under-fitting, additional optimization of the hyper-parameters is likely possible. 

To train the NNP for the coarse-grained MD with LiCl, we extract the coordinates and forces for the ions alone for 24,000 frames extracted from a 2.4 ns NNP-MD all-atom simulation. A larger dataset is required to sufficiently converge the averaging over the solvent degrees of freedom. It would not be feasible to generate such a large training dataset with FPMD directly, but it is straightforward with NNP-MD. The coarse-grained NNP requires many fewer weights and trains very quickly in comparison to the all-atom NNP due to the much simpler energy surface, we found that reducing the number of parameters was important to provide stability, meaning less physical inaccuracies in the simulation.  Two layers of interaction blocks were used with the maximum $l$ set to 1 each with 8 features. Only even parity was used. Invariant neurons for the radial network were set to 8. All the other parameters were set to the defaults. The radial cutoff was also extended to 10 \AA \ to provide longer-range interactions. 

Figure~\ref{fig:LCs} shows the learning curves. The RMSE on the validation set for the all-atom NNPs was between 9-12 meV/\AA\ for the forces and 0.1-0.14 meV for the energies. 

Figure~\ref{fig:ForceErrorHistogram} shows a histogram of errors for the all atom LiCl NNP, the linear correlation plot is shown in Figure~\ref{fig:Forcecorre}.

The RMSE on the ions is much higher with the coarse-grained model, as expected, due to the neglect of the solvent. They were 299 meV/\AA\ for the forces and 38 meV for the energies for LiCl. 

The error on initialisation was 302 meV/\AA\ and 91 meV, respectively, meaning that training of the NNP only removed 3 meV/\AA\ in error on the forces, yet this is enough to reliably reproduce the ion-ion RDFs surprisingly.

For the KCl coarse-grained force field, a higher concentration (6.6 M) was used, and the errors were 228 meV/\AA\ and 88 meV. Fewer frames (7500) were needed compared with LiCl (24,000) due to the higher concentration. 

To test the all atom NNP could simulated KCl at high concentrations, 261 frames were extracted from the 6.6 M run and resampled with DC-r$^2$SCAN. These were used to train an additional all-atom neural network potential, which showed good agreement with the results with the model trained on data at 2.5 M. (See Figure~\ref{fig:RDFKCll6p6M-HCtraincomp})

\subsection{NNP MD (NequIP/LAMMPS)}
The NequIP  plugin for LAMMPS\cite{Plimpton1995}  was used to perform several NVT simulations at 300 K for over a nanosecond each. A Nos\'e-Hoover thermostat was attached to every degree of freedom to ensure equilibration~\cite{Martyna1992}.  The long-range Coulomb interactions were added to the simulation using the LAMMPS hybrid overlay method.  No initial data was discarded, as the initial frame was taken from the end of the FPMD simulation or classical simulation.

The 2.5 M simulations in the larger cell size contained 48 ions and 512 water molecules. The 1.3 M simulations contained 24 ions and 512 water molecules.

For LiCl, the 2.5 M cell had dimensions of 25.26$^3$ \AA$^3$ and the 1.3 M simulations had a 25.05$^3$ \AA$^3$ cell size. For KCl, a 25.46$^3$ \AA$^3$ box was used with the same composition. For LiBr, a 25.39$^3$ \AA$^3$ box was used with the same composition. For the KCl at 6.6 M a smaller box size of 14.47$^3$ \AA$^3$ was used as long range electrostatic interactions are likely to be well screened at such a high concentrations.  The cell sizes were computed to match the experimental density. 

The total simulation times were 8.4 ns for the 2.5 M LiCl; 12 ns for the 1.3 M LiCl; 5.3 ns for the 2.5 M KCl; 3.5 ns for the 6.6 M KCl; 2.7 ns for LiBr 2.5 M.  

VMD\cite{spivakVMDPlatformInteractive2023} was used to create the RDFs, images, and videos.

The infinite dilution PMFs were computed in LAMMPS by simply computing the total energy of the system of two ions in a large box as a function of distance. At short distances, where the NNP has no data as the ions do not approach closely, the NNP can oscillate randomly. At these points, an increasing extrapolation was used to ensure that the  infinite dilution PMF didn't go negative again. 

We also include a small Gaussian repulsion in the OH interaction to prevent occasional hydrogen dissociation from water. This only acts in the 0 density region so should have no effect on the simulations.  

The coarse grained NNPs were also used for molecular dynamics simulations with identical parameters to the all atom MD simulations. These were run for 6 million steps (3 ns).  

The KCl slab simulation contained 216 KCl inintialised in a crystal structure, in a 18.62\AA x 18.62\AA 28.62\AA  cell, where there was a 10 \AA vacuum region. 
  
\subsection{Kirkwood-Buff theory calculations}
Kirkwood-Buff theory \cite{Kusalik1987} was used to compute the derivatives of the activities from the RDFs using the following expressions: 
\begin{equation}
\frac{d\ln\gamma}{d\ln c}=\frac{1}{1+\rho \left(G_{\text{cc}}-G_{\text{co}}\right)}
\end{equation}
where $\rho$ is the ion density. For monovalent ions,  $G_{\text{cc}}$ is given by:
\begin{equation}
G_{\text{cc}}=\frac{1}{4}\left(G_{++}+G_{--}+2G_{+-}\right)
\end{equation}
and $G_{\text{co}}$ is given by:
\begin{equation}
G_{\text{co}}=\frac{1}{2}\left(G_{+\text{O}} + G_{-\text{O}}\right)
\end{equation}
where $G$ refers to the Kirkwood-Buff integrals:
\begin{equation}
G_{ij} = \int_0^\infty \left( g_{ij}(r) - 1 \right) r^2 \, dr
\end{equation}
The integrals were cutoff at half the box size and the RDFs were normalised to ensure the average value around $\pm$ 2 \AA \ of the cutoff went to 1. 

\subsection{Diffusion coefficients calculation}
Diffusion coefficients were computed from the mean squared displacements (MSD) of the water molecules and lithium and chloride ions in our NNP MD trajectories. This conversion was carried out using the diffusion coefficient-MSD relationship described below:
\begin{equation}
    D = \frac{\text{MSD}}{6t}
\end{equation}
where the MSD is:
\begin{equation}
\text{MSD}(t) = \frac{1}{N} \sum_{i=1}^{N} \left| \mathbf{r}_i(t) - \mathbf{r}_i(0) \right|^2
\end{equation}
and $t$ is time.
The results were finally adjusted by finite size corrections. (See eq.~12 of Ref.~\citenum{Yeh2004}.) Here, we have used the experimental value (0.888 mPas) for the viscosity of pure water when determining the finite size correction. Experimental values were obtained from Ref.~\citenum{tanakaMeasurementsTracerDiffusion1987}.


\subsection{Classical force field}
The classical molecular dynamic simulation was performed for NaCl using the LAMMPS program, applying the Dang-Smith\cite{Smith1994} Lennard-Jones force field. Cross interactions were computed using the Lorentz-Berthelot mixing rule. A 5 ns simulation was conducted with an NVT ensemble at 300 K, controlled by the Nose-Hover thermostat. The simulation was carried out in a box with a volume of 29.8$^3$ \AA$^3$, using a periodic boundary conditions. The system contained 64 sodium ions, 64 chloride ions and 810 SPC/E water molecules, representing a 4M NaCl solution. Long-range Coulombic interactions were calculated by particle-particle particle-mesh method with a relative force set as 10$^ {-5}$. The cut-off was set 15 \AA\ , slightly extended to improve accuracy in long-range Coulombic interactions.

The classical model was used to provide additional testing of the coarse-grained model. The process is similar to that described for first principle coarse-grained model. Initially, the training data was generated from a 5 ns classical run of a 4M NaCl solution by rerunning the classical trajectory and extracting only ions data every 20 frames, resulting in a total of 25000 frames. Again the long-range electrostatic interactions were removed from forces and energies in the dataset.
In NNP training, all the hyperparameters were set as previously mentioned, except for a longer radial cutoff of 15 \AA\ and higher maximum $l$ of 2. Finally, we simulated NaCl at 1.3M, 2.6M and 4M using thbe coarse-grained MD simulations with the force field trained on the classical MD simulation of NaCl at 4M. All other hyper parameters for the CG-NNP trained on CMD simulations were the same as the hyperparameters for the CG-NNP trained on all atom NNP-MD. 

\section{Acknowledgements} 
We acknowledge Stefan Vuckovic, Debra Bernhardt, Christopher Mundy, Gregory Schenter, Simon Batzner, Alby Musaelian, James Stevenson, Sophie Baker, Alister Page, Juerg Hutter and Andrey Bliznyuk for helpful discussions and computational assistance. TTD was supported by an Australian Research Council (ARC) Discovery Project (DP200102573) and DECRA Fellowship (DE200100794). TG was supported by an Australian Research Council (ARC) Discovery Project (DP200100033) and Future Fellowship (FT210100663). This research was undertaken with the assistance of resources and services from the National Computational Infrastructure (NCI), which is supported by the Australian Government and with the assistance of resources from QCIF. (http://www.qcif.edu.au)
\bibliography{Libraray-zotero}
\clearpage
\section{Supplementary Information}

\begin{suppfigure}[htbp]
 \begin{subfigure}[]{0.49\textwidth}
      	 \includegraphics[width=\textwidth]{{{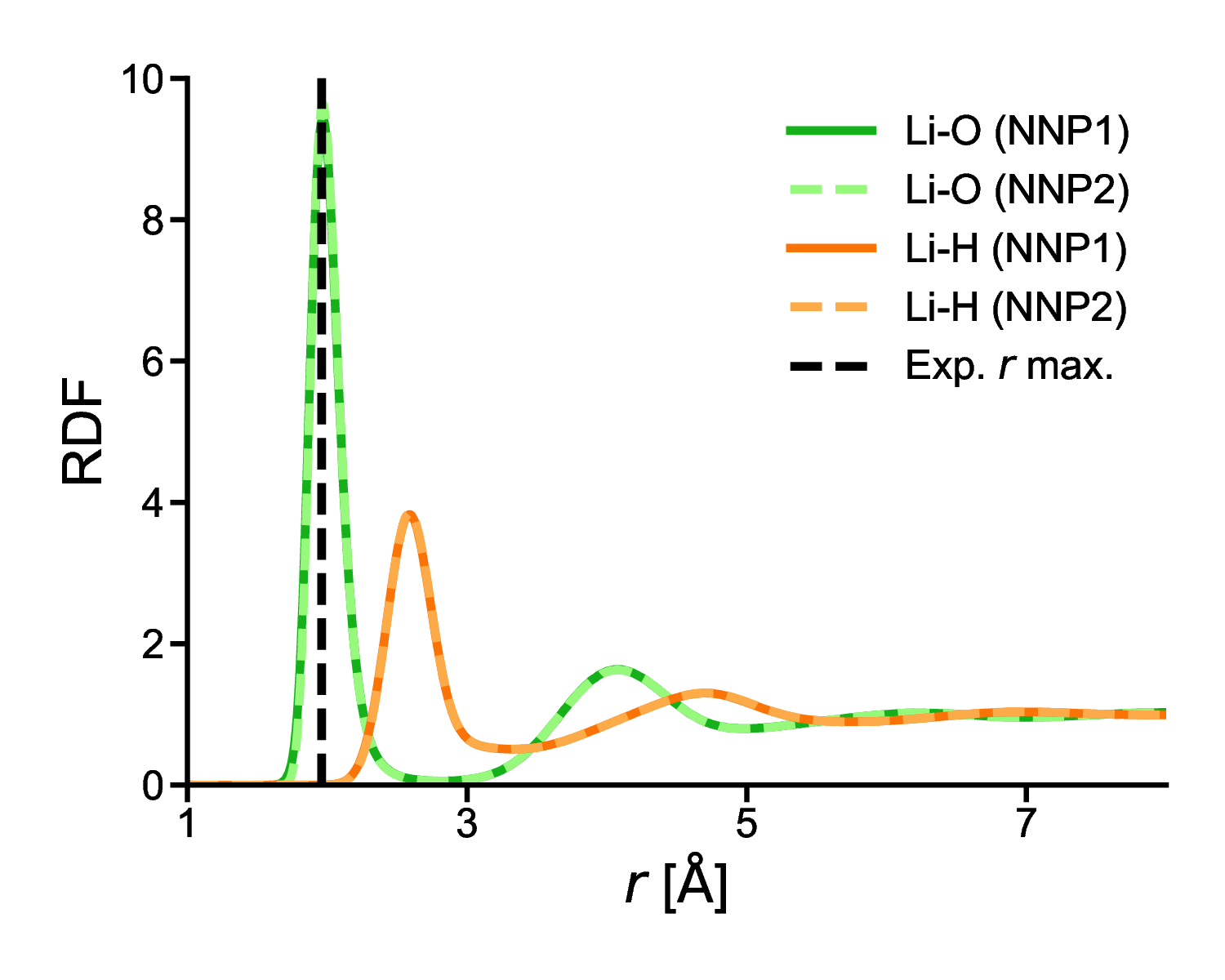}}}
        \caption[]{Lithium water RDFs in LiCl.}
        \label{fig:RDFLiwatcomp}
\end{subfigure}
 \begin{subfigure}[]{0.49\textwidth}
      	 \includegraphics[width=\textwidth]{{{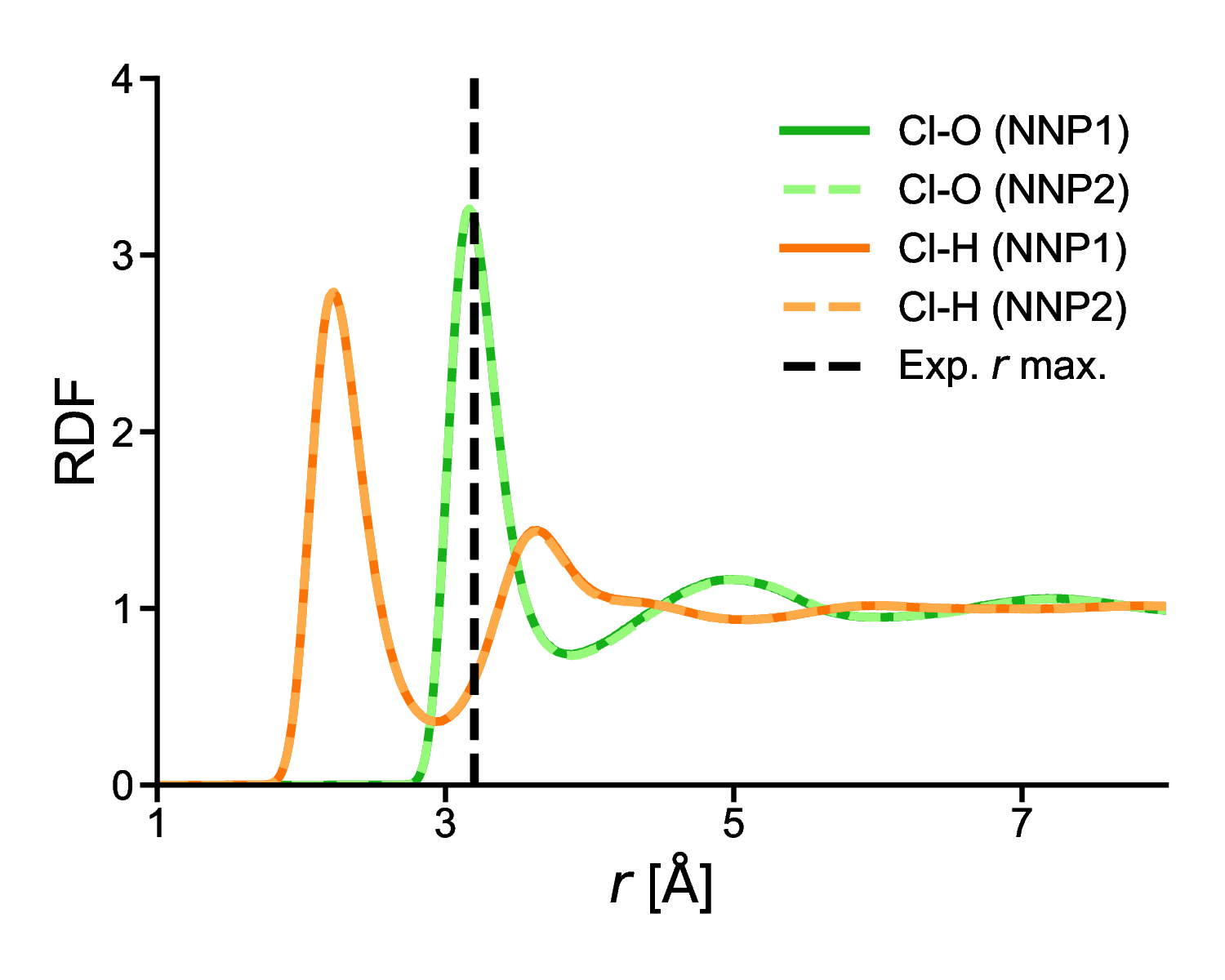}}}
        \caption[]{Chloride water RDFs in LiCl}
        \label{fig:RDFClwatcomp}
\end{subfigure}
\\ 
\begin{subfigure}[]{0.49\textwidth}
      	 \includegraphics[width=\textwidth]{{{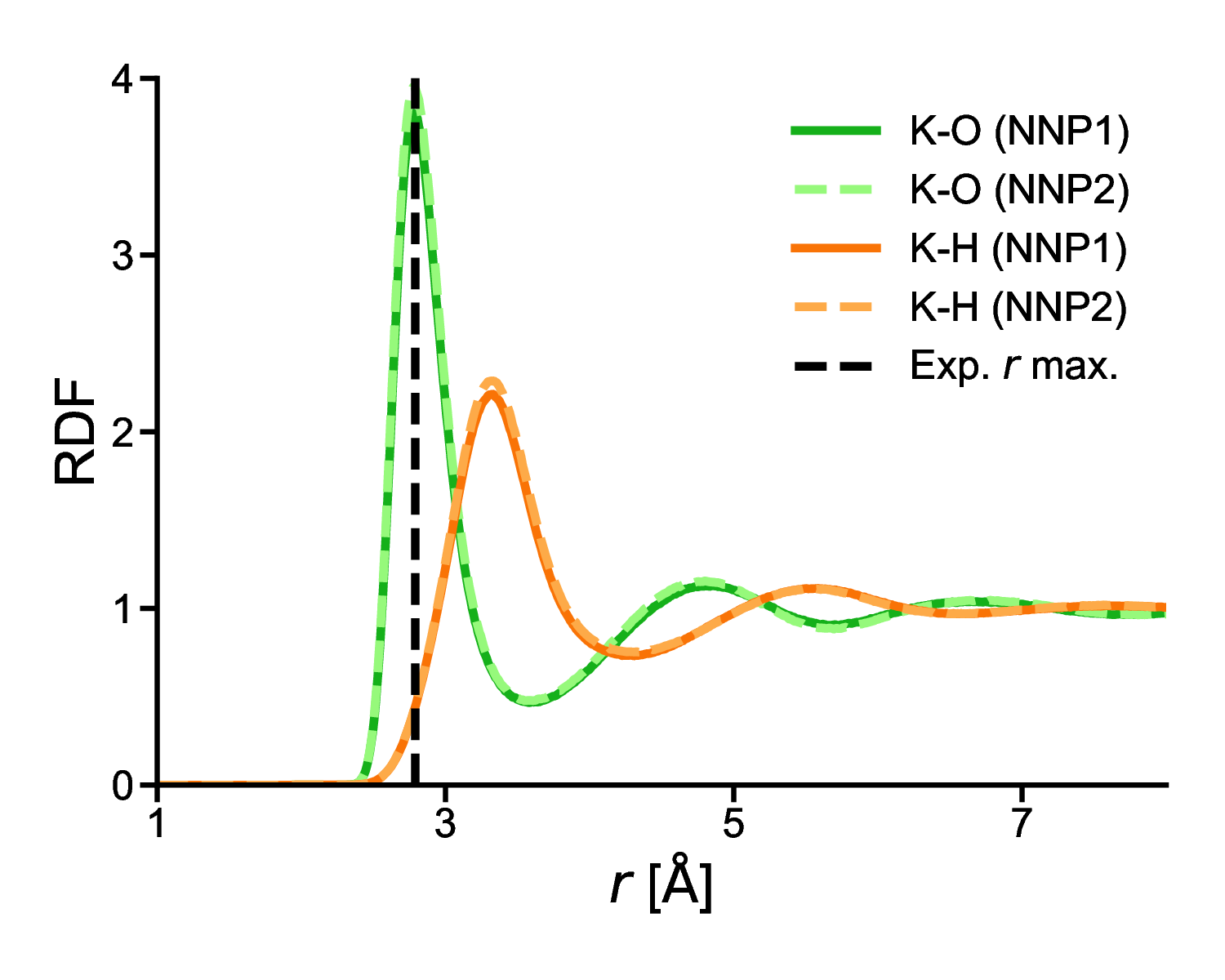}}}
        \caption[]{Potassium water RDFs.}
        \label{fig:RDFKwatcomp}
\end{subfigure}
\begin{subfigure}[]{0.49\textwidth}
      	 \includegraphics[width=\textwidth]{{{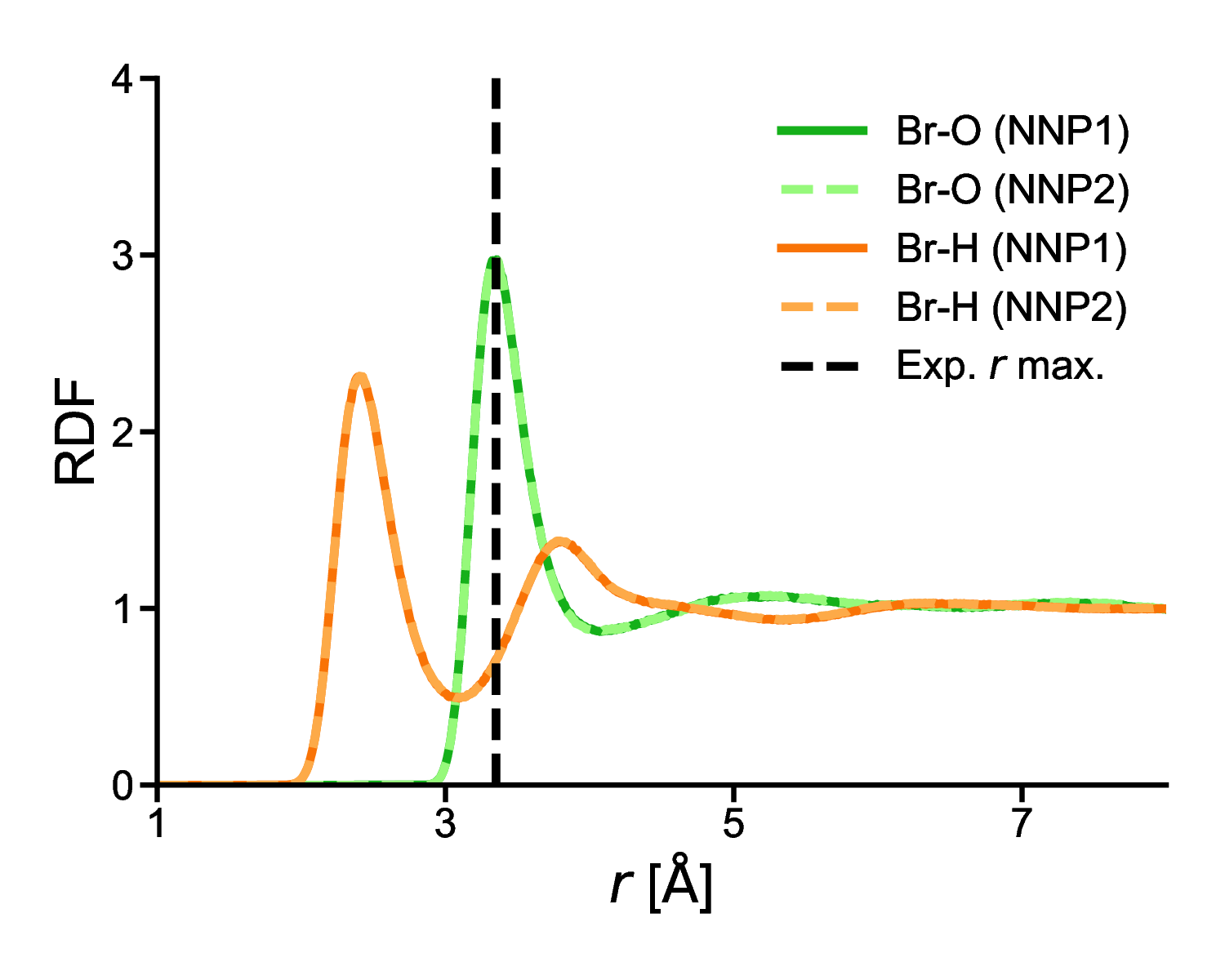}}}
        \caption[]{Bromide water RDFs}
        \label{fig:RDFClLiBrwatcomp}
\end{subfigure}
\caption{Ion water RDFs at 2.5 M, with the all atom NNP-MD. }
\label{fig:RDFIonwatcomp}
\end{suppfigure}

\begin{suppfigure}[htbp]
\centering
\includegraphics[width=.7\textwidth]{{{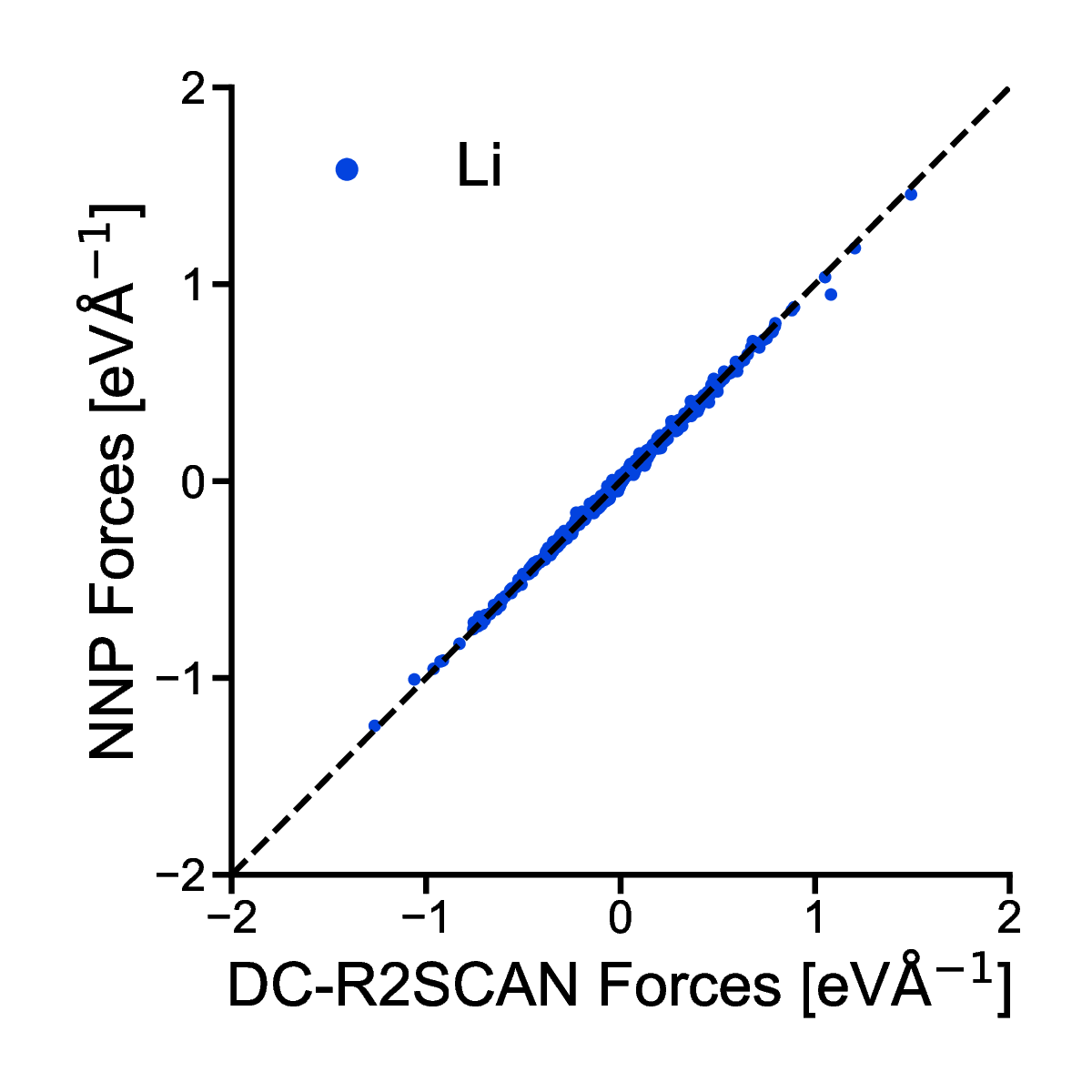}}}
\caption{Forces on the lithium ions in the lithium dimer computed with DFT compared with NNP predictions showing good agreement.}\label{fig:ForcecorrelLidimer}
\end{suppfigure}

\begin{suppfigure}[htbp]
  \begin{subfigure}[]{0.49\textwidth}
      	 \includegraphics[width=\textwidth]{{{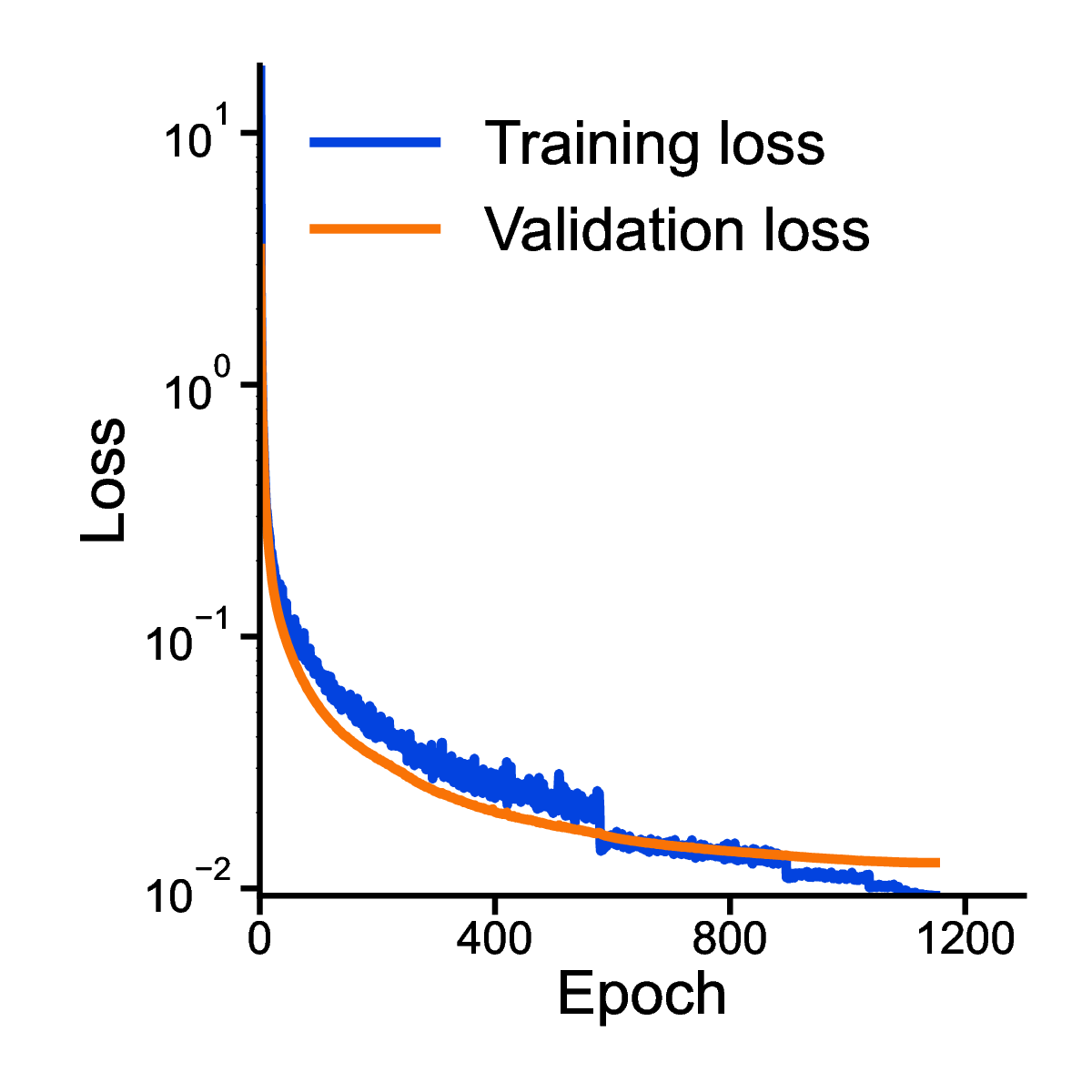}}}
       \caption[]{All-atom LiCl NNP}
        \label{fig:LCl3}
\end{subfigure}
 \begin{subfigure}[]{0.49\textwidth}
      	 \includegraphics[width=\textwidth]{{{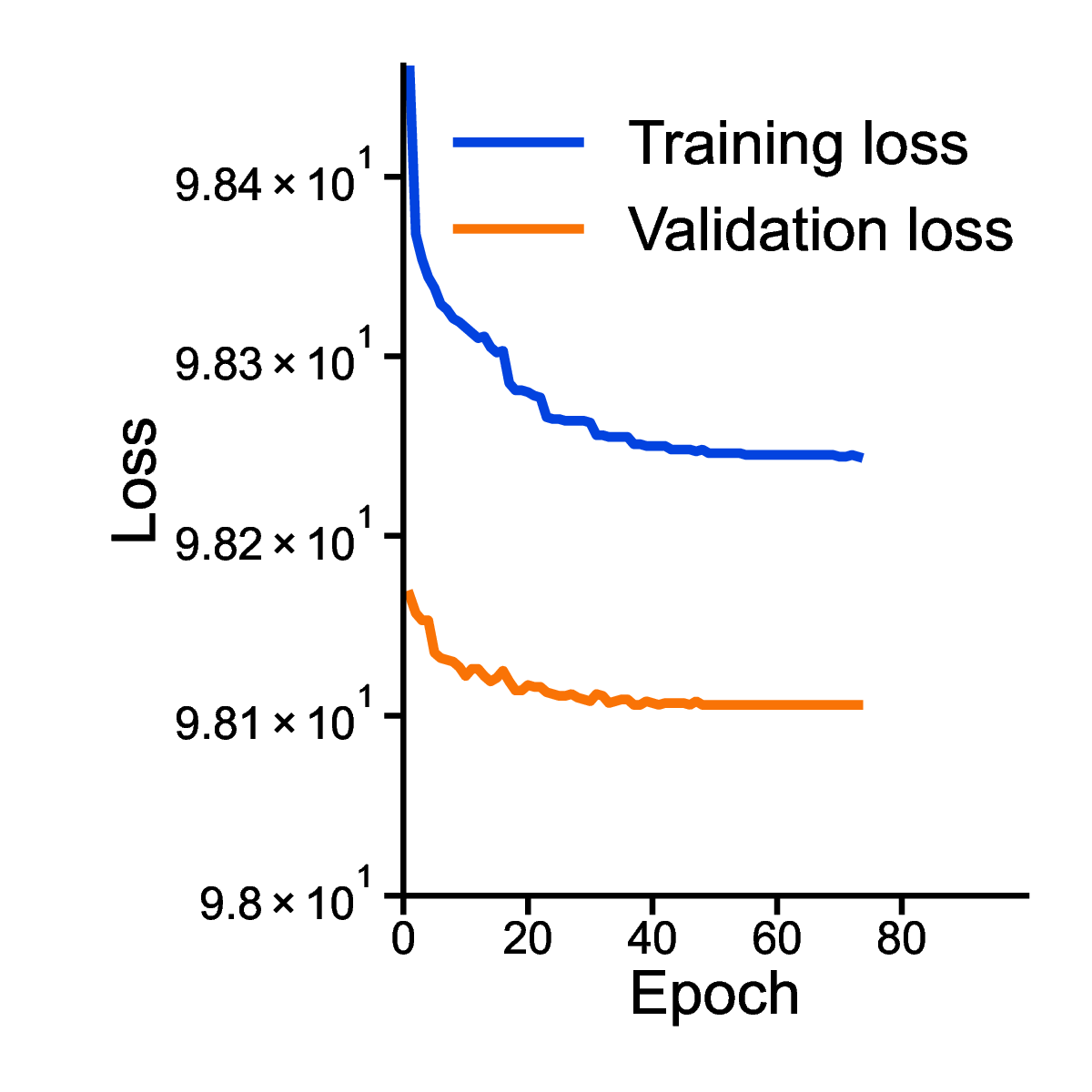}}}
        \caption[]{Coarse-grained LiCl NNP}
        \label{fig:LCl3-cg15}
\end{subfigure}
\\ \begin{subfigure}[]{0.49\textwidth}
      	 \includegraphics[width=\textwidth]{{{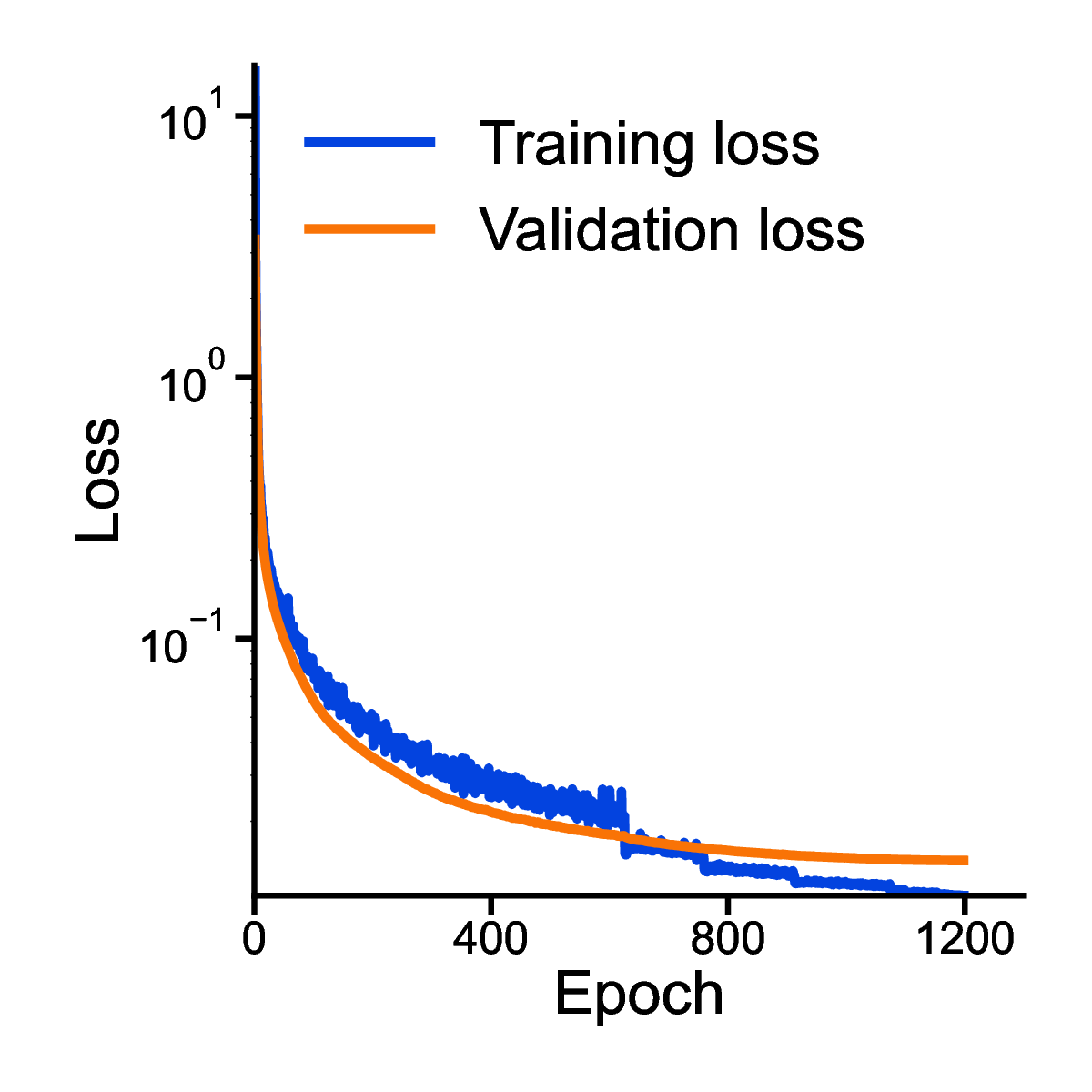}}}
        \caption[]{All-atom KCl NNP}
        \label{fig:LCkc3}
\end{subfigure}
 \begin{subfigure}[]{0.49\textwidth}
\includegraphics[width=\textwidth]{{{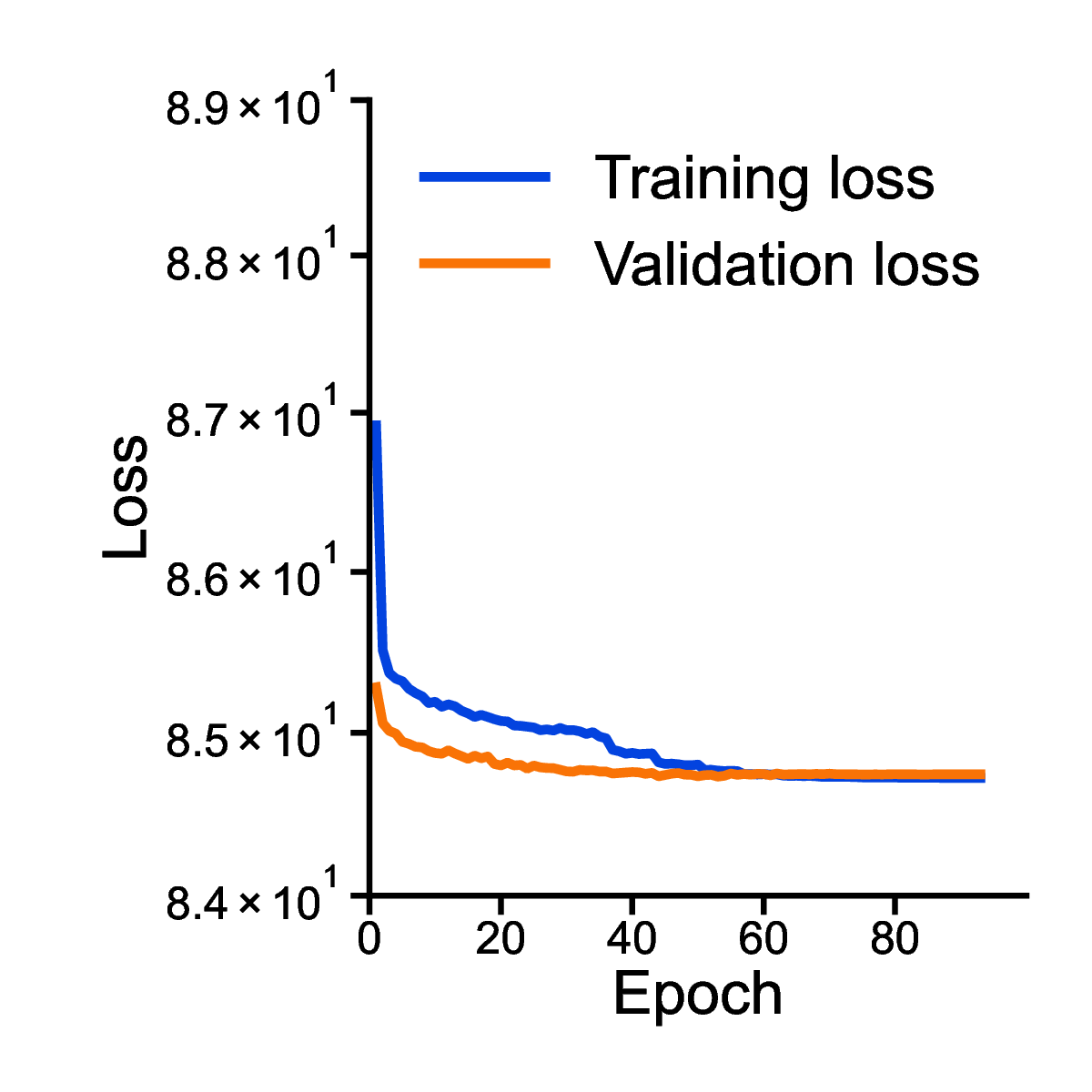}}}
        \caption[]{Coarse-grained KCl NNP}
        \label{fig:LCkc3-6p6M-cg1}
\end{subfigure}
\\
 \begin{subfigure}[]{0.49\textwidth}
\includegraphics[width=\textwidth]{{{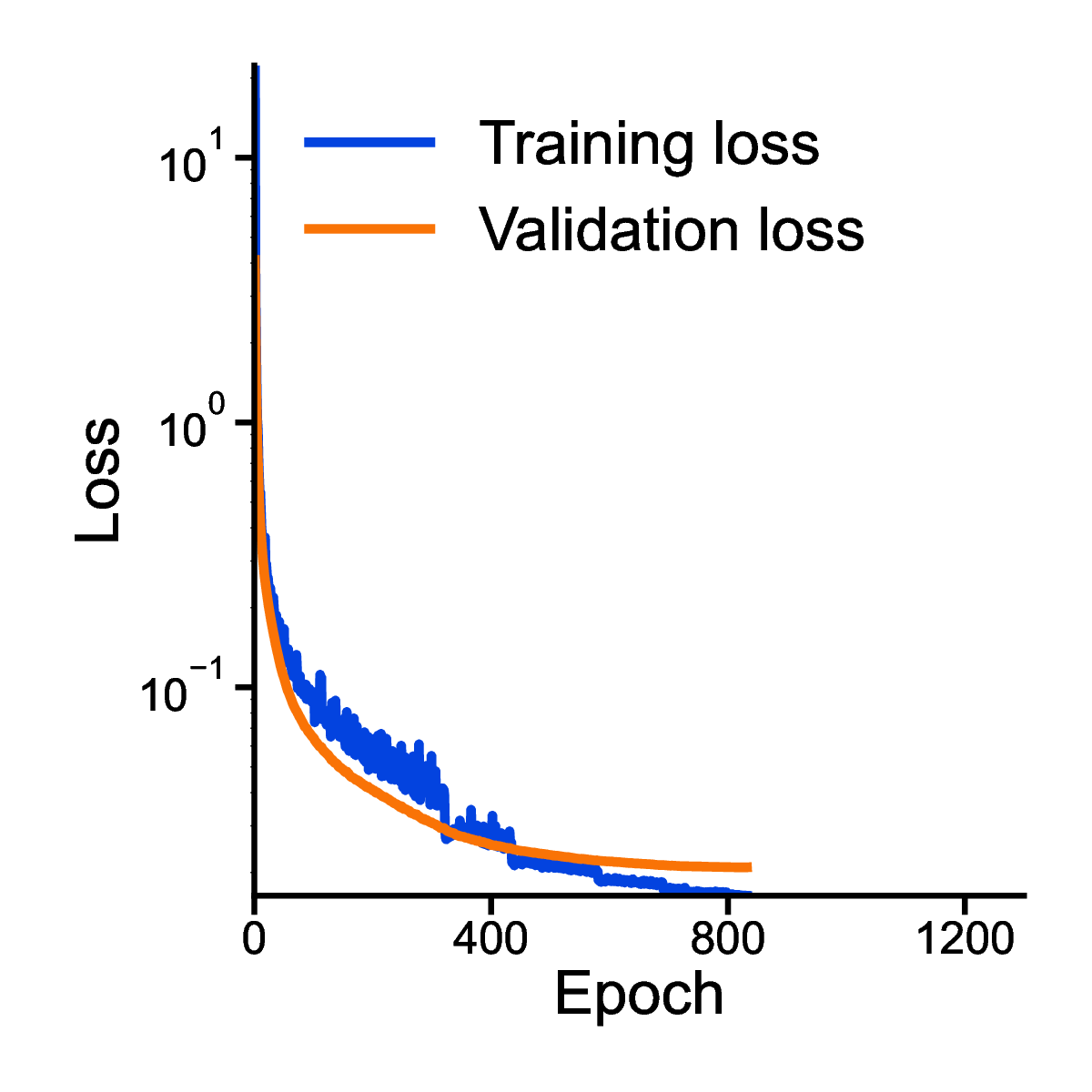}}}
        \caption[]{All-atom LiBr NNP}
        \label{fig:LClb3}
\end{subfigure}
\caption{Learning curves for all-atom and coarse-grained NNPs.}
\label{fig:LCs}
\end{suppfigure}

\begin{suppfigure}[htbp]
    \centering
    \includegraphics[width=.7\textwidth]{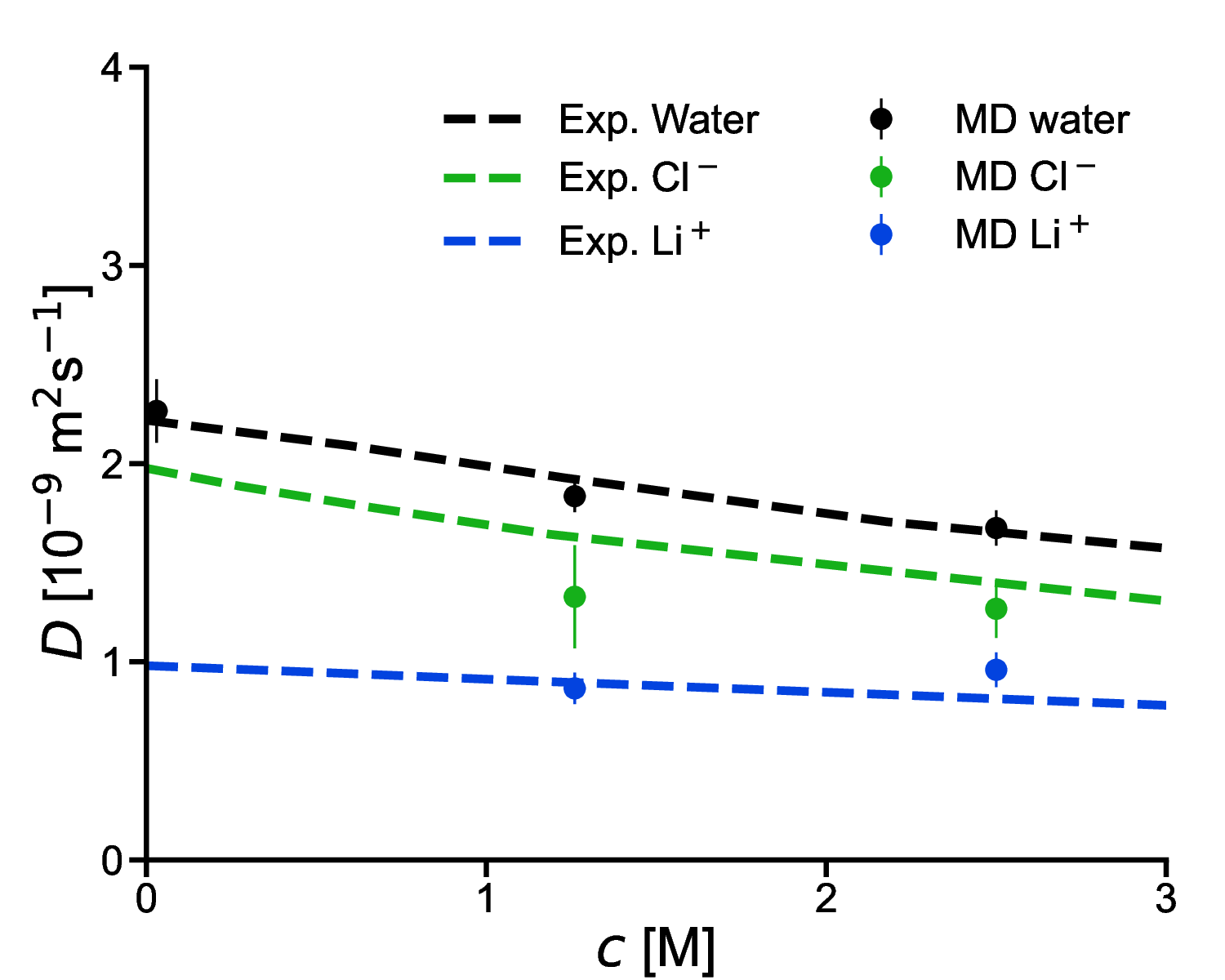}
    \suppfigcaption{Comparison of the computed diffusivities with experimental values\cite{tanakaMeasurementsTracerDiffusion1987} for water and both ions at two concentrations and infinite dilution.}
      \label{fig:Diffs}
\end{suppfigure}

\begin{suppfigure}[htbp]
\centering
\includegraphics[width=.7\textwidth]{{{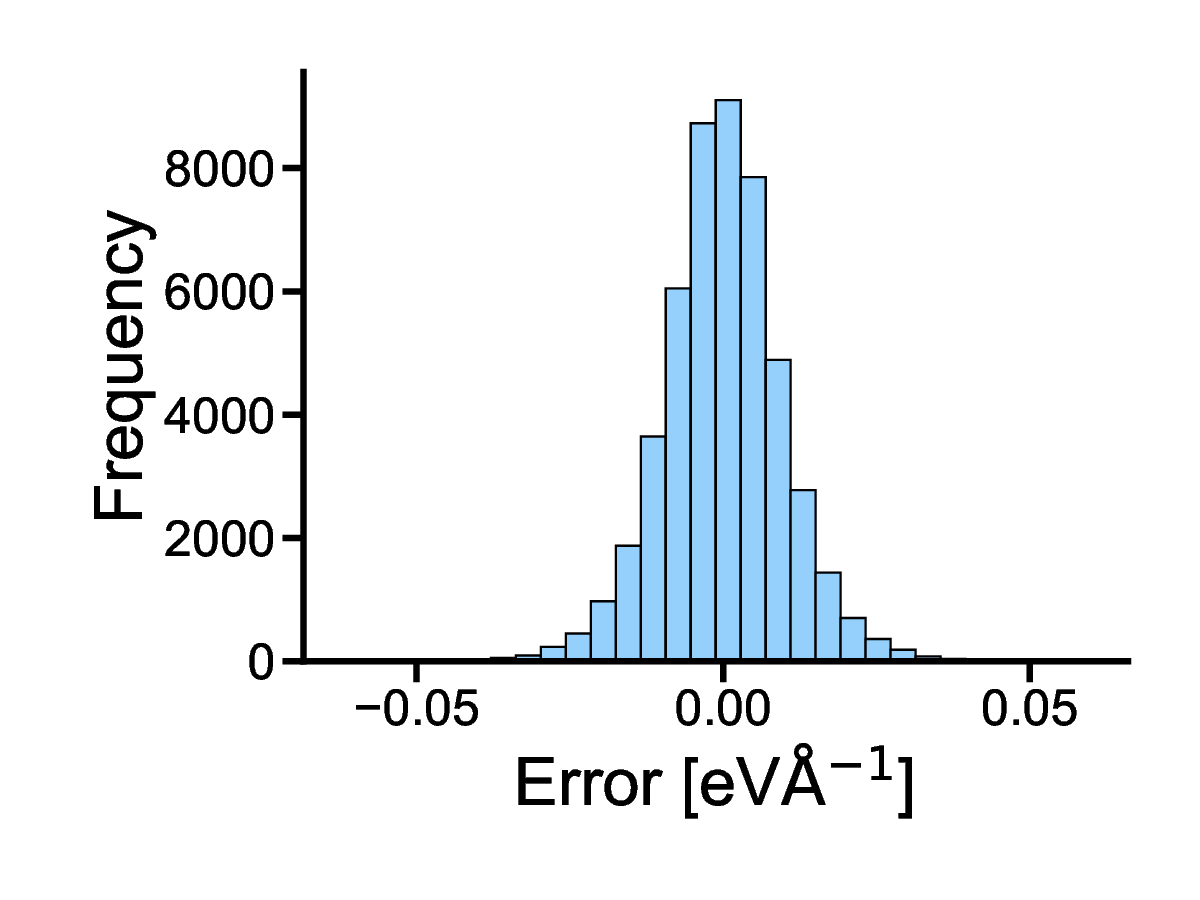}}}
\suppfigcaption{Histogram of errors for the all atom LiCl NNP. Linear correlation plots are shown in Figure~\ref{fig:Forcecorre}}\label{fig:ForceErrorHistogram}
\end{suppfigure}

\begin{suppfigure}[htbp]
\begin{subfigure}[]{0.49\textwidth}
      	 \includegraphics[width=\textwidth]{{{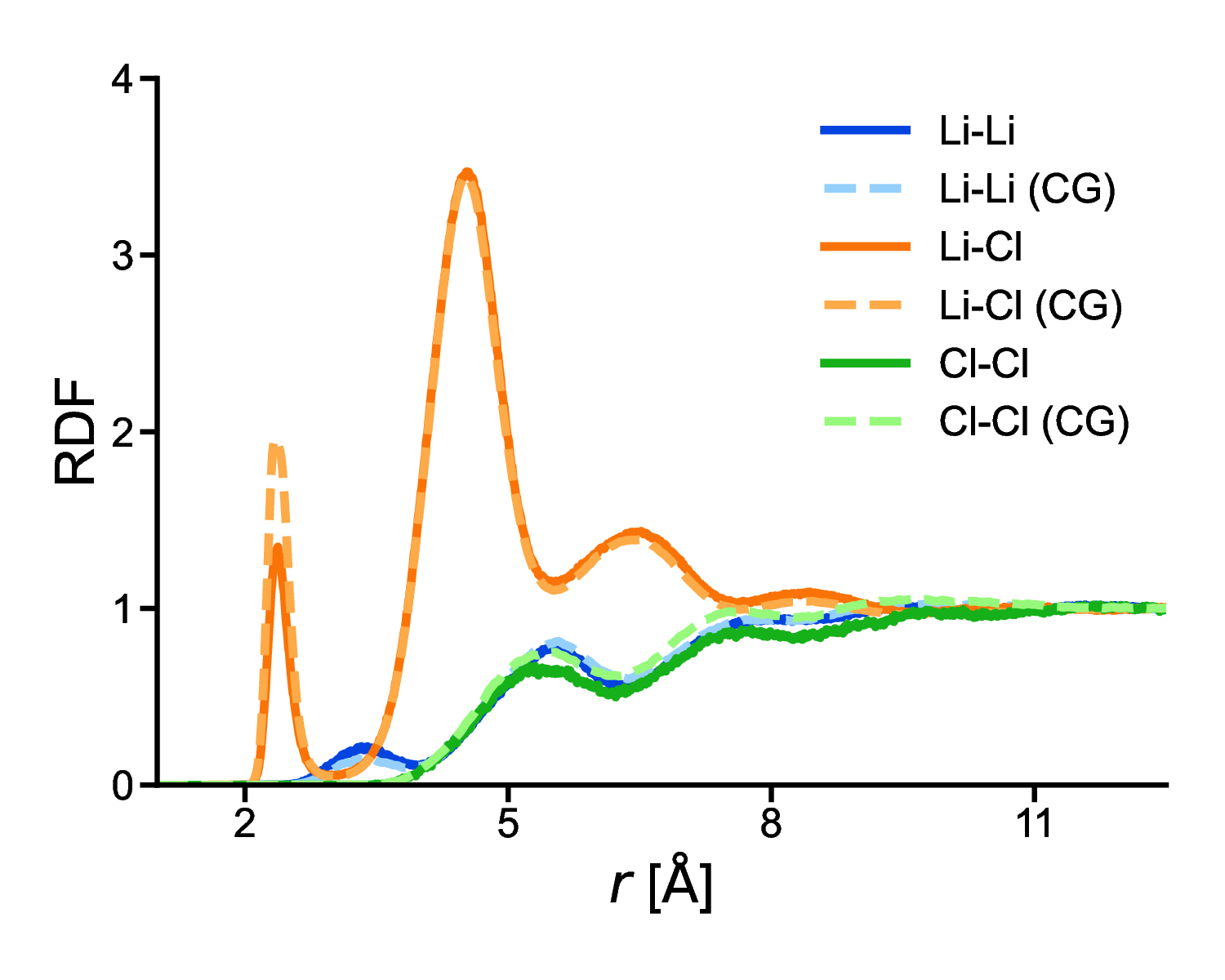}}}
        \caption[]{LiCl at 1.3 M.}
        \label{fig:RDFLiCll31p2M-cg15comp}
\end{subfigure}
  \begin{subfigure}[]{0.49\textwidth}
      	 \includegraphics[width=\textwidth]{{{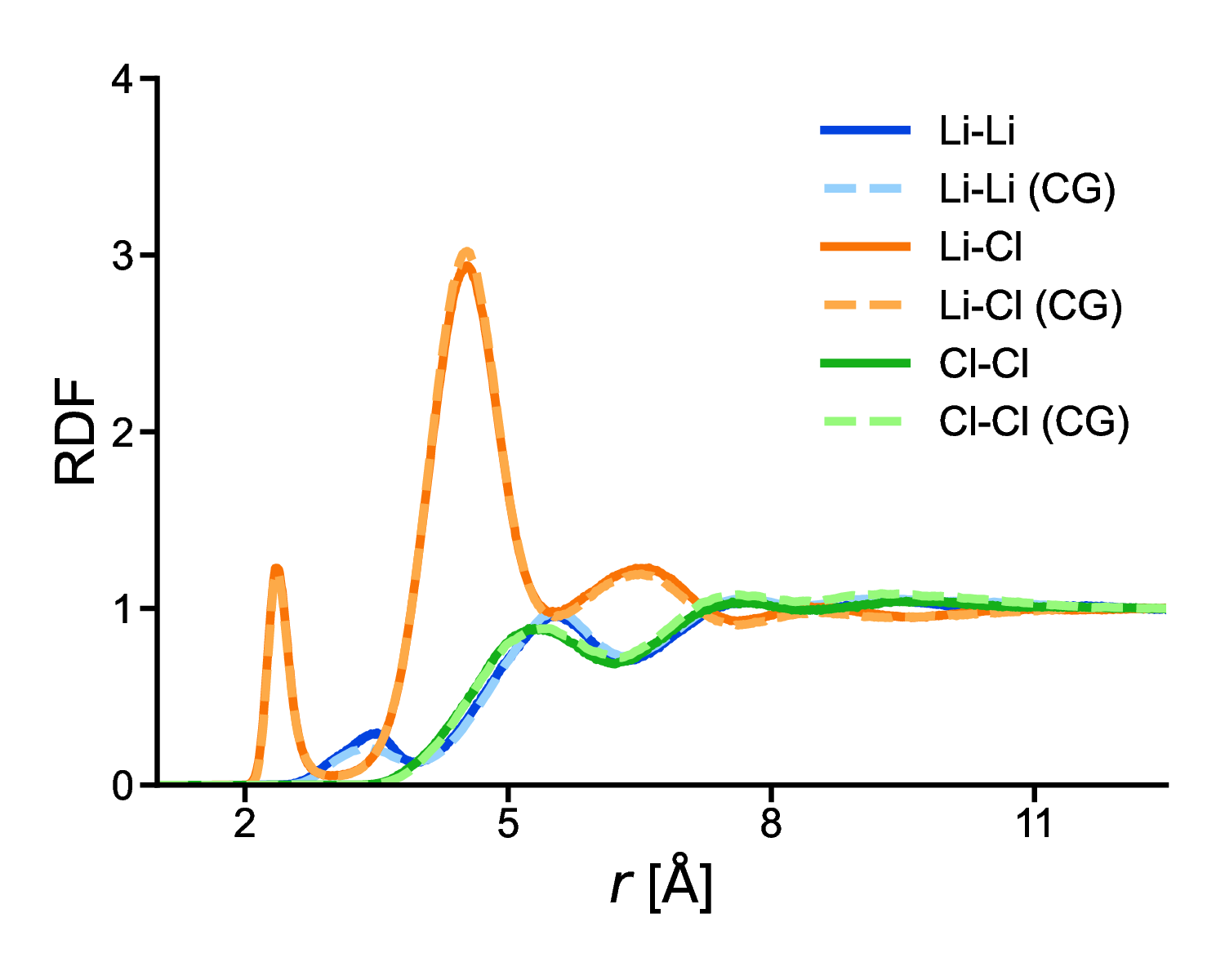}}}
        \caption[]{LiCl at 2.5 M.}
        \label{fig:RDFLiCll32p4M-cg15comp}
\end{subfigure}
\caption{Comparison of RDFs computed with all atom NNP-MD and with with coarse-grained NNP-MD for (a) LiCl at 1.3 M and (b) LiCl at 2.5 M}
 \label{fig:cgcomp}
\end{suppfigure} 

To validate that the NNP can extrapolate to higher concentrations we simulated KCl at 6.6 M with an NNP trained on data at 6.6 M and one trained at 2.5 M. Good agreement is shown in Figure~\ref{fig:RDFKCll6p6M-HCtraincomp}.

\begin{suppfigure}[htbp]
      	 \includegraphics[width=\textwidth]{{{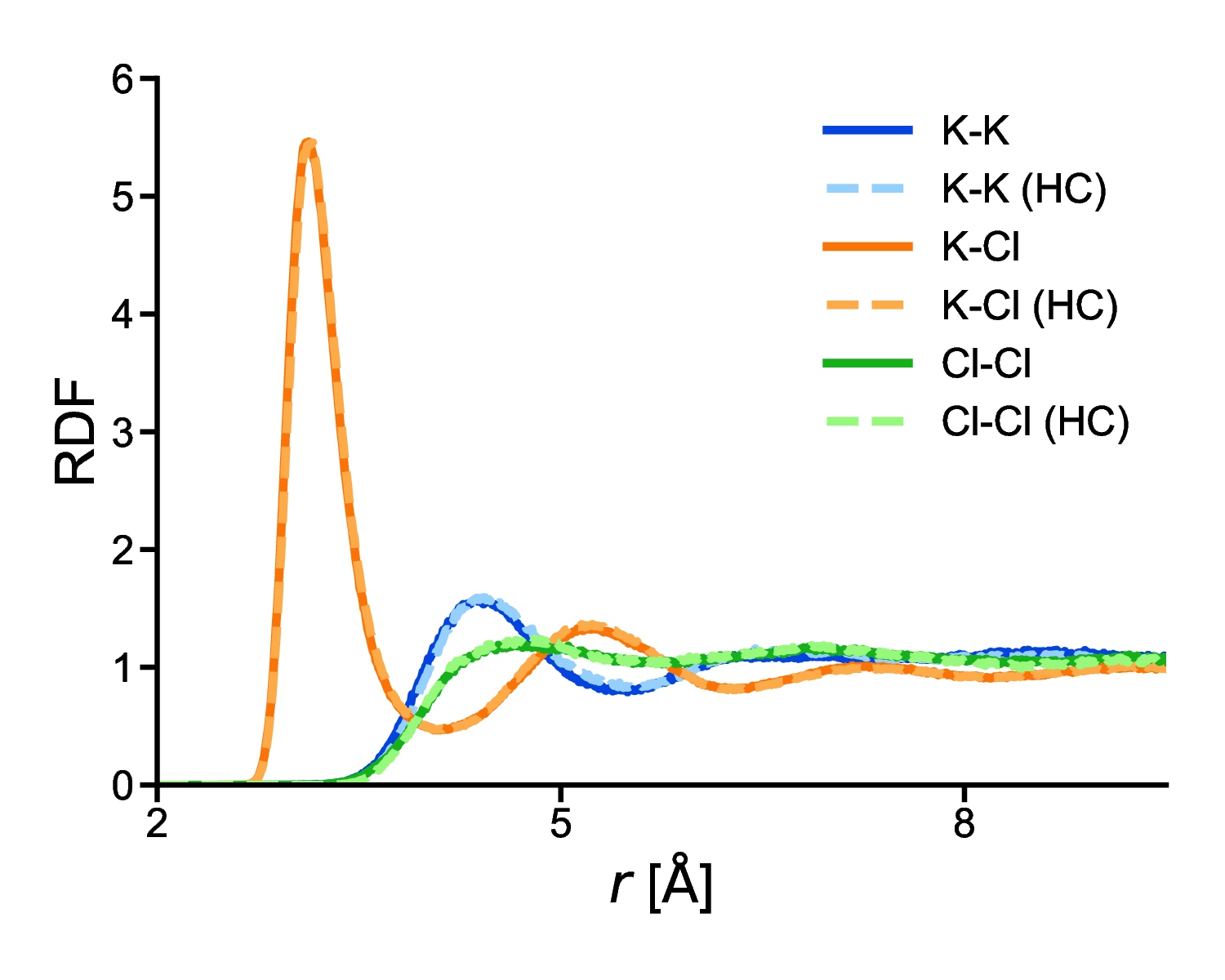}}}
        \caption[]{Ion-ion RDFs for KCl at 6.6 M with NNP trained at 2.5 M vs 6.6 M (HC) }
        \label{fig:RDFKCll6p6M-HCtraincomp}
 \label{fig:cgcompHC}
\end{suppfigure} 

\begin{suppfigure}[htbp]
  \begin{subfigure}[]{0.49\textwidth}
      	 \includegraphics[width=\textwidth]{{{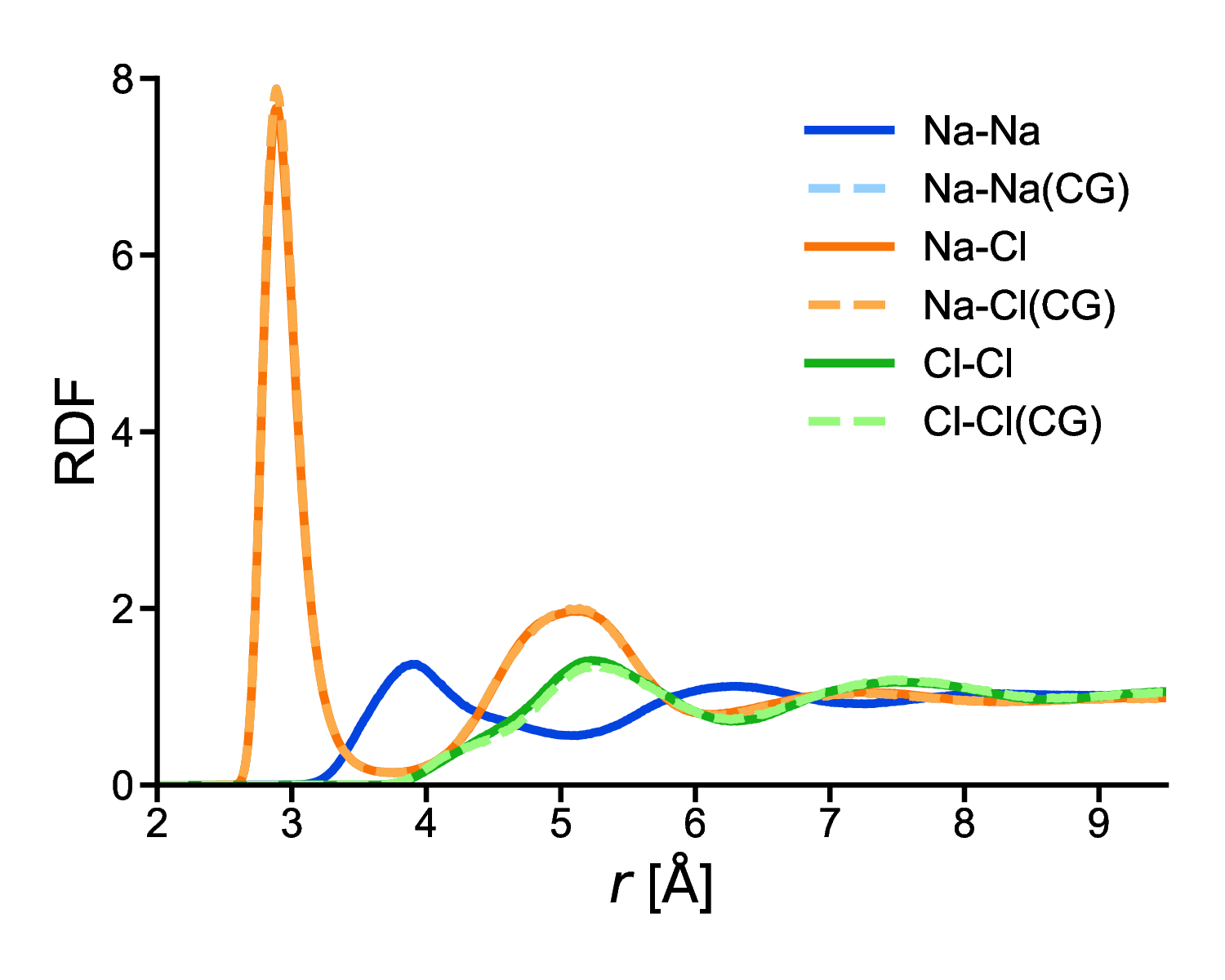}}}
        \caption[]{4M}
        \label{fig:pythonRDF4MCMDvsCGMD}
\end{subfigure}
\begin{subfigure}[]{0.49\textwidth}
      	 \includegraphics[width=\textwidth]{{{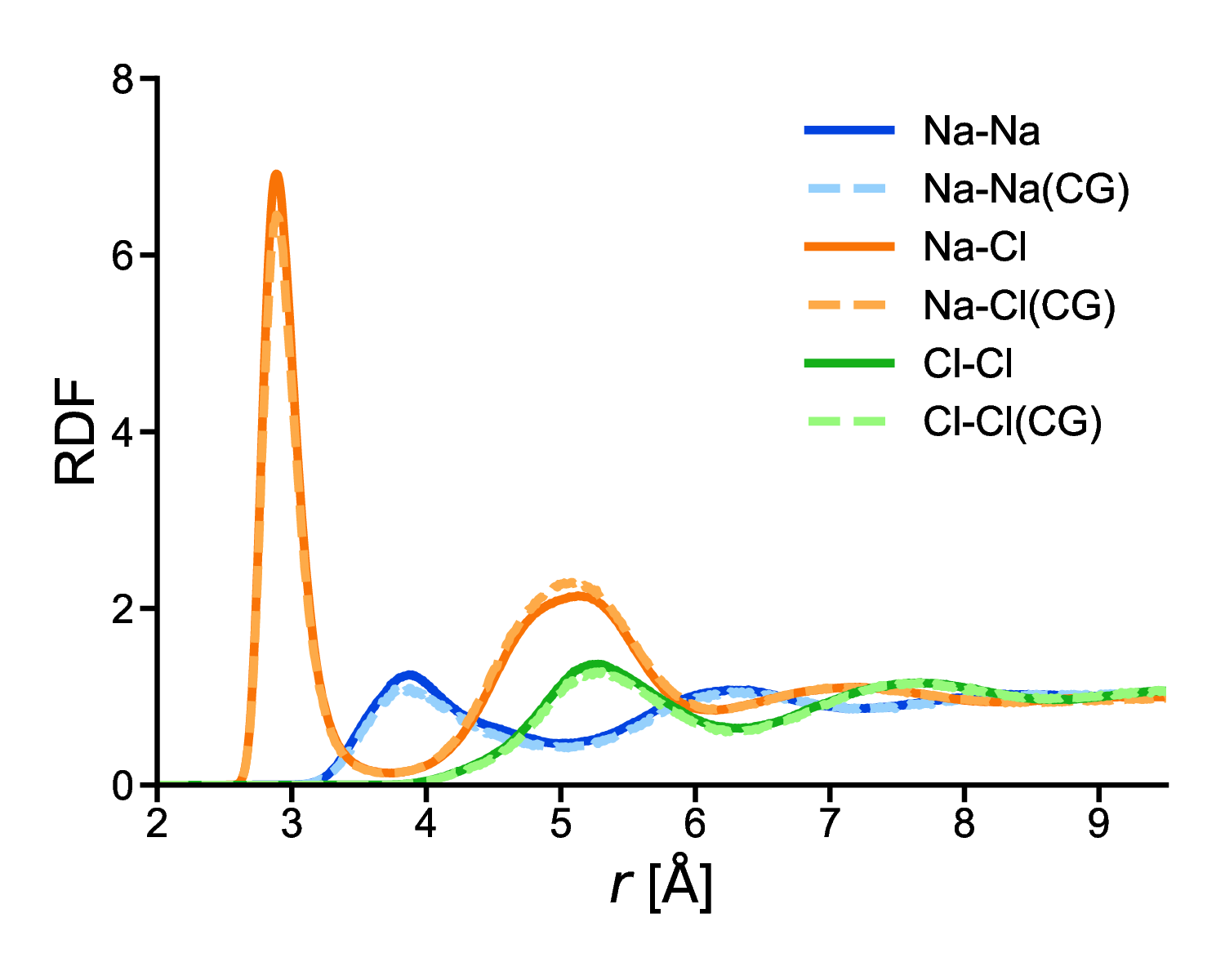}}}
        \caption[]{2.6 M}
        \label{fig:pythonRDF2p6MCMDvsCGMD}
\end{subfigure}
\begin{subfigure}[]{0.49\textwidth}
      	 \includegraphics[width=\textwidth]{{{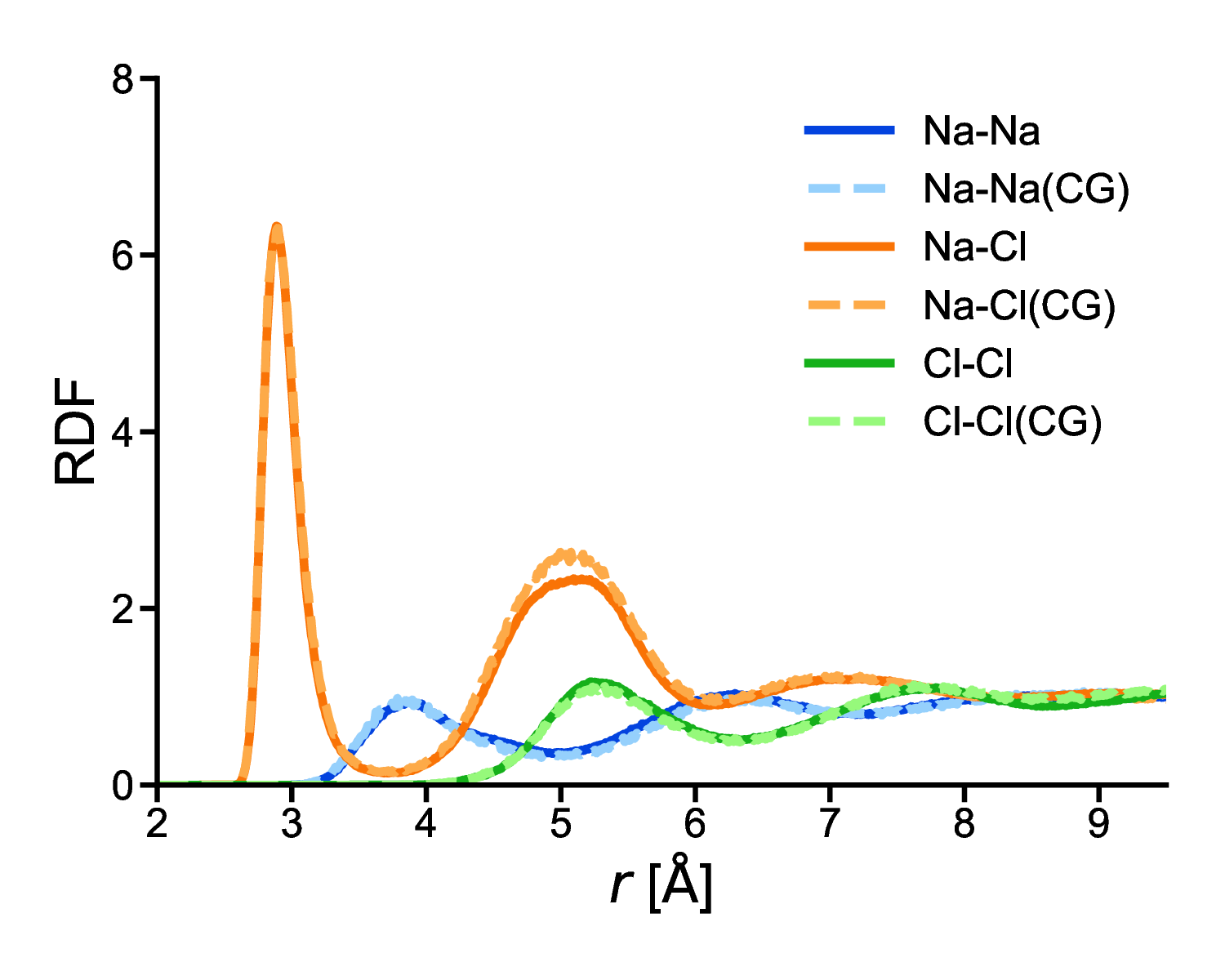}}}
         \caption[]{1.3M}
        \label{fig:pythonRDF1p3MCMDvsCGMD}
\end{subfigure}
\caption{Comparison of RDFs computed with all atom NNP-MD and with with coarse-grained NNP-MD for .}
\label{fig:RDFCMDvsCGMD}
\end{suppfigure}

\subsection{Ion water RDFs}
The full ion water RDFs are shown in Figure~\ref{fig:RDFIonwatcomp}. 

\subsection{Water-water RDFs}
Oxygen hydrogen and hydrogen-hydrogen RDFs are shown in Figure~\ref{fig:OHHHRDFs}.

\begin{suppfigure}[htbp]
  \begin{subfigure}[]{0.49\textwidth}
      	 \includegraphics[width=\textwidth]{{{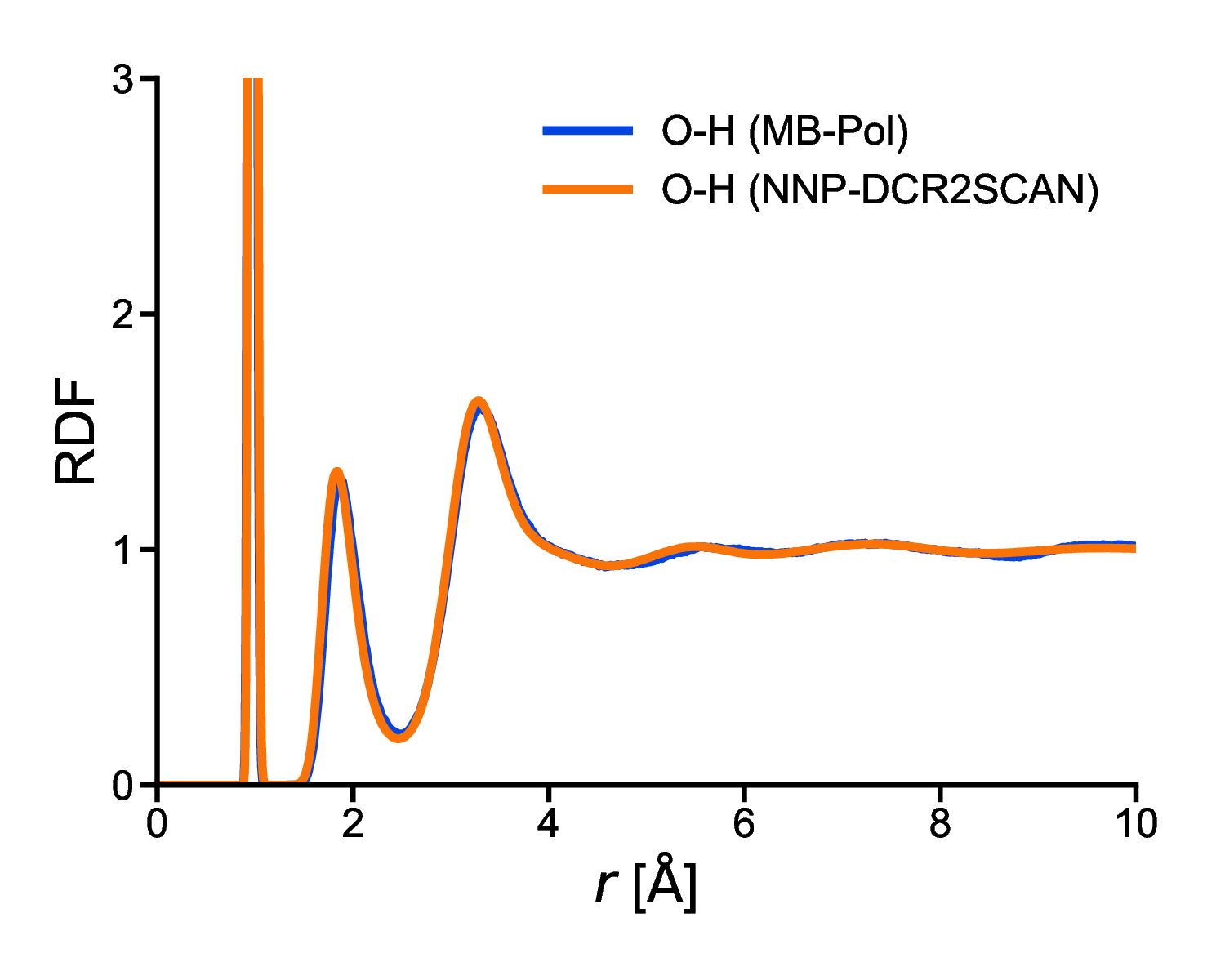}}}
        \caption[]{Oxygen-hydrogen RDF}
        \label{fig:OHRDF}
\end{subfigure}
\begin{subfigure}[]{0.49\textwidth}
      	 \includegraphics[width=\textwidth]{{{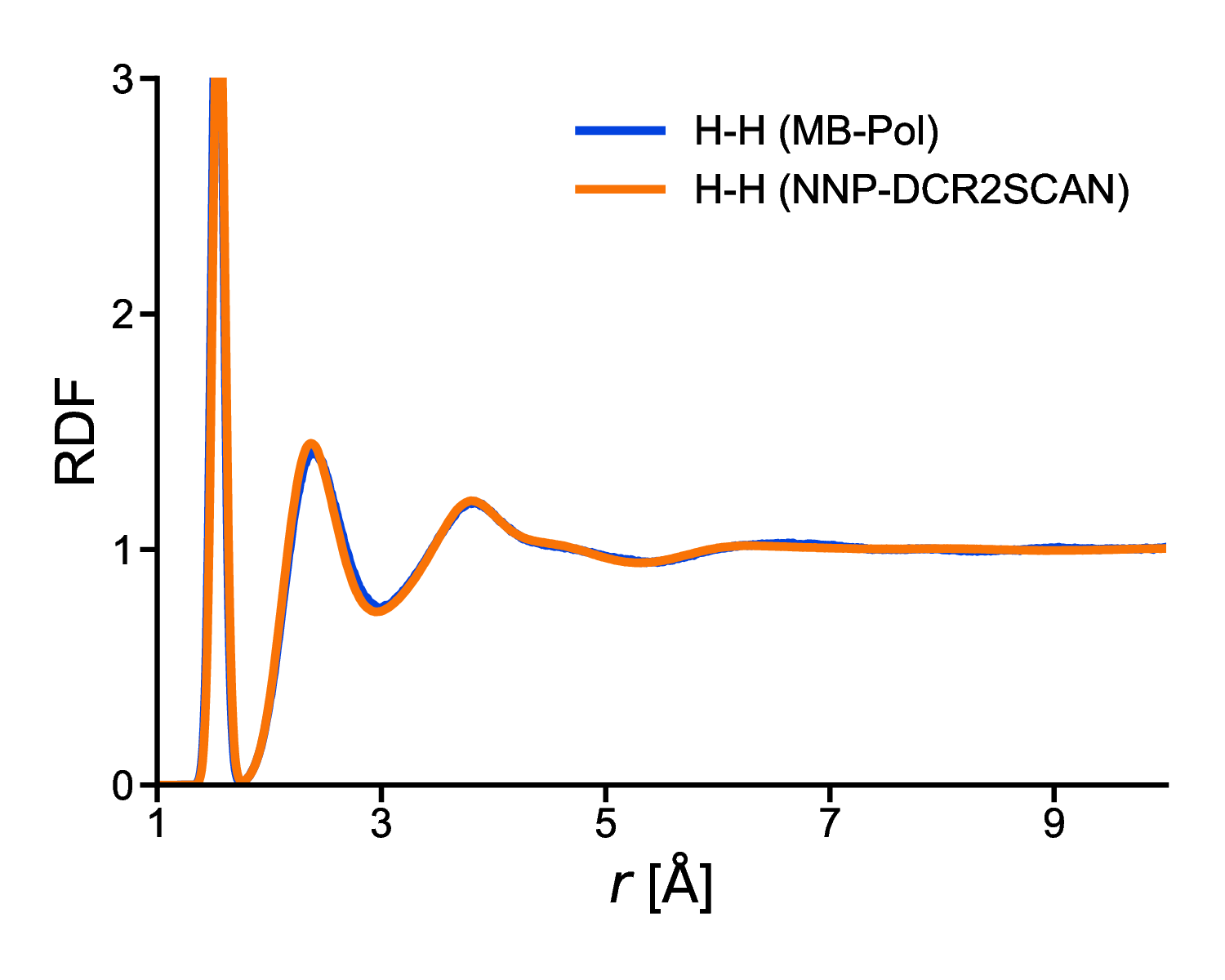}}}
        \caption[]{Hydrogen-hydrogen RDF}
        \label{fig:HHRDF}
\end{subfigure}
\caption{Oxygen-hydrogen and hydrogen-hydrogen RDFs with the NNP trained on 2.4 M LiCl in comparison with MB-Pol.}
\label{fig:OHHHRDFs}
\end{suppfigure}

\subsection{Error metrics}

\begin{table}
    \centering
    \begin{tabular}{c|cc}
        Model &  Force RMSE (meV/\AA) &Energy RMSE (meV) \\
        LiCl (NNP1) & 8.9 &0.11 \\
        LiCl (NNP2) & 9.7 &0.12 \\
        KCl (NNP1) &9.5 &     0.11   \\
        KCl (NNP2) &  9.6& 0.11 \\
        LiBr (NNP1) &  11.7 &     0.14   \\
        LiBr (NNP2) & 10.7 & 0.16\\
        LiCl-CG &  299  &  38.0  \\
        KCl-CG (6.6M) & 228 &   80.0 \\
        NaCl-CG (4M) & 311 &   15.7 \\
    \end{tabular}
    \caption{Validation metrics for the various NNPs trained. LiCl-CG was trained on all atom MD with LiCl (NNP1) at 2.5 M and KCl-CG was trained on all atom MD with KCl (NNP1) at 6.6 M.  NaCl-CG was trained using all atom CMD simulations at 4M. The RMSE forces are component wise averages over the different species.}
    \label{tab:my_label}
\end{table}
 
\subsection{Lithium dimer forces}
To confirm that the lithium dimer formed is physically reasonable we recomputed the forces with DC-DFT on the lithium ions in the dimer conformation and compared with the NNP predictions acheiving good agreement (Figure~\ref{fig:ForcecorrelLidimer}).

\subsection{Learning curves}
Learning curves for all the models trained are shown in Figure~\ref{fig:LCs}. The very low error of the all atom models is evidence combined with the higher error but quick convergence of the coarse grained models. 

\subsection{Diffusivities}
The diffusivities as a function of concentration for Li, Cl and water are given in Figure~\ref{fig:Diffs}. Again good agreement is observed including the gradual decease in diffusivity of water as a function of ion concentration. 

\subsection{Coarse grained models}
A comparison of the coarse grained and all atom models for LiCl at 2.5 M and 1.3 M are shown in Figure~\ref{fig:cgcomp}.
Showing that the decrease in Debye length as a function of concentration is well reproduced. Similarly. the coarse grain model for NaCl trained on 4 M simulations is shown to be able to predict RDFs at different concentrations in Figure~\ref{fig:RDFCMDvsCGMD}.






\end{document}